\documentclass[sn-apa]{sn-jnl}% APA Reference Style 

\usepackage{graphicx}%
\usepackage{multirow}%
\usepackage{amsmath,amssymb,amsfonts}%
\usepackage{amsthm}%
\usepackage{mathrsfs}%
\usepackage{array}
\usepackage[title]{appendix}%
\usepackage{xcolor}%
\usepackage{textcomp}%
\usepackage{manyfoot}%
\usepackage{booktabs}%
\usepackage{algorithm}%
\usepackage{algorithmicx}%
\usepackage{algpseudocode}%
\usepackage{listings}%
\usepackage{longtable}
\usepackage{tabularx}
\usepackage[utf8]{inputenc}
\usepackage[style=apa, backend=biber]{biblatex}
\begin{document}

\title[Article Title]{Phase model analysis of the effect of M-current on neural synchrony in hippocampal networks}

\author*[1]{\fnm{Megha} \sur{Manoj}}\email{ms2manoj@uwaterloo.ca}

\author*[1]{\fnm{Sue Ann} \sur{Campbell}}\email{sacampbell@uwaterloo.ca}

\affil*[1]{\orgdiv{Department of Applied Mathematics and Centre for Theoretical Neuroscience}, \orgname{University of Waterloo}, \orgaddress{\city{Waterloo}, \postcode{N2L 3G1}, \state{ON}, \country{Canada}}}

\maketitle
\begin{abstract}

Neural assemblies, transiently coordinated groups of neurons, observed in the hippocampus are thought to underlie the formation of episodic memories. Acetylcholine (ACh), a neuromodulator, that is received by the hippocampus, plays a critical role in memory and learning. A well-supported hypothesis suggests that high levels of ACh during active exploration and rapid eye movement (REM) sleep promote memory encoding, while low levels during quiet waking and slow-wave sleep (SWS) support memory consolidation. We study this bidirectional role of ACh in neural assembly formation through its effect on the synchrony among neurons. We consider a network model of pyramidal neurons, each equipped with a slow, voltage-dependent, non-inactivating potassium current (M-current), which is downregulated in the presence of ACh. Neural assemblies are represented as cluster solutions to this system. Using a one-dimensional phase model
reduction of a pair of weakly coupled pyramidal neurons under different levels of the M-current, we predict the symmetric cluster solutions that may emerge in larger networks equipped with all-to-all globally homogeneous, symmetric distance-dependent and nearest-neighbours coupling architectures. We find that under low ACh conditions, the network can fully synchronize, whereas high levels can desynchronize the network into multiple stable symmetric cluster solutions representing distinct neural assemblies.
\end{abstract}
\section{Introduction}
The study of synchronization within neural networks is crucial for understanding how different regions of the brain communicate and coordinate on both local and global scales. In particular, groups of neurons may transiently synchronize to produce spatiotemporal patterns of activity, which serve as a mechanism for encoding information. A network of spiking neurons can organize into such distinct patterns by forming ‘neural assemblies’, which are groups of neurons with  transiently synchronized activity. 

 Neural assemblies observed in the hippocampus in particular, are thought to facilitate spatial navigation tasks and the encoding and storage of episodic memories (autobiographical memories) \parencite{byrne_molecules_2014,  harris_organization_2003,dragoi_temporal_2006,pastalkova_internally_2008,sakurai_hippocampal_1996}.  It is also theorized that, in addition to external inputs, the intrinsic oscillatory dynamics of the hippocampal neurons can play a role in the formation of neural assemblies \parencite{harris_organization_2003,pastalkova_internally_2008} .  Acetylcholine is a  neuromodulator that affects hippocampal activity and is known for its vital role in cognitive functions, particularly those involving learning and memory \parencite{hasselmo_modes_2011,hasselmo_dynamics_1995,hasselmo_laminar_1994}.  Different levels of ACh are theorized to be linked to distinct neural activity that emerge during different physiological states (wakefulness and sleep)  \parencite{huang_acetylcholine_2022}.  A generally accepted consensus, is that high levels of ACh during active exploration and rapid eye movement (REM) sleep, support the encoding of new stimuli, while low levels of ACh during quiet waking and slow wave sleep (SWS), facilitate the consolidation of previously formed memories for future recall \parencite{hasselmo_dynamics_1995,hasselmo_encoding_1996,hasselmo_high_2004, hasselmo_neuromodulation_1999,huang_acetylcholine_2022,gais_low_2004}. In the present article, we investigate this bidirectional role of acetylcholine (ACh) in neural assembly formation, specifically studying the distinct neural assemblies that may be supported by a network of pyramidal neurons in the hippocampus, under varying levels of ACh (high vs. low).
 
 Mathematically, neural assemblies are reminiscent of cluster solutions observed in biophysical models of neuronal networks, and thus are often suggested as a framework for studying assembly formation \parencite{galan_predicting_2006,kilpatrick2011sparse,ryu_spatially_2021,miller_patterns_2022}. A cluster solution is a special type of phase-locked solution of the network, where neurons self-organize into groups, called clusters. A zero phase difference is maintained among the neurons in a given cluster, while neurons in different clusters maintain fixed nonzero phase differences. 
 % synchronous clusters.
 In particular, by imposing symmetries in the coupling architecture of the network, we can study model-independent cluster solutions, where the inter-cluster phase differences are identical. The resulting solutions are symmetric, in the sense that each cluster contains the same number of neural oscillators. To study these types of solutions, we derive a phase model approximation of the biophysical network, an approach that has been used previously to investigate cluster solutions in networks of weakly coupled oscillators \parencite{okuda_variety_1993,ermentrout_frequency_1984,li_clustering_2003,li_clustering_2003-1,miller_clustering_2015,campbell_phase_2018,ryu_spatially_2021,miller_patterns_2022}. A phase model approximation can simplify a high-dimensional biophysical neuron model to a simpler model while retaining essential dynamical features sufficient to study neural synchrony \parencite{galan_efficient_2005}. By assuming circulant coupling,  the corresponding biophysical network satisfies periodic boundary conditions, effectively arranging the identical model neurons along a one-dimensional ring. Phase model analysis has been used to derive the conditions of existence and stability of symmetric cluster solutions for such networks under various coupling architectures, including all-to-all coupling with synaptic weights that are a globally homogeneous  \parencite{okuda_variety_1993,campbell_phase_2018}, or have a symmetric heterogeneous distribution \parencite{campbell_phase_2018,miller_clustering_2015,ryu_spatially_2021}.  We apply these results to a homogeneous network of conductance based model pyramidal neurons coupled via weak excitatory synapses, under varying levels of ACh, and consider three coupling configurations: all-to-all globally homogeneous, symmetric distance-dependent and nearest-neighbours coupling.   

Acetylcholine acts as a neuromodulator by altering the neuronal behaviour at both the cellular and network levels. At the cellular level,  acetylcholine can suppress \parencite{halliwell_voltage-clamp_1982} a slow, voltage-dependent, non-inactivating potassium current 
found in hippocampal cells \parencite{brown_neural_2009}  known as the muscarinic current (M-current). Our work is primarily inspired by the results of a previous paper \parencite{al-darabsah_m-current_2021}, where  an M-current induced Bogdanov Takens bifurcation in a general conductance based neuron model with an M-current is linked to a switch in the neuronal excitability from class I to class II.  They applied their results to a model for a pair of hippocampal pyramidal neurons coupled with an excitatory \textit{AMPA} synapse and demonstrated an M-current induced transition from anti-phase solutions (class I excitability, no M-current) to in-phase/synchronized (class II excitability, high M-current), through intermediary out-of-phase solutions (intermediate strengths of M-current). This result aligns with prior studies of pairs of model neurons weakly coupled together with an excitatory synapse and either class I or class II excitability \parencite{hansel_synchrony_1995,ermentrout_type_1996}.  The M-current induced switch in the neuronal excitability, has been extensively studied, both experimentally and computationally. Experimental studies have demonstrated changes in the shape and type of the phase response curves (PRC) associated with cortical pyramidal neurons, reflecting the changes in their excitability, primarily driven by the M-current, which have then been simulated for different neuron models \parencite{stiefel_cholinergic_2003,stiefel_cholinergic_2008,stiefel_effects_2009}. Computational papers have also studied specific network level implications of the M-current induced switch in neuronal dynamics \parencite{hu_two_2002,fink_cellularly-driven_2011,zhou_m-current_2018}. 
However, to our knowledge, these works do not systematically address the manner in which the hippocampal network undergoes desynchronization, when the M-current is suppressed, as it would in the presence of acetylcholine. In \parencite{roach_acetylcholine_2019} and \parencite{roach_formation_2015}, they identify two information coding regimes associated with high and low levels of ACh respectively via numerical simulations. In particular, both papers observed a higher degree of synchronization within the network in the presence of M-current.  In \parencite{ermentrout_effects_2001}, they demonstrate a desynchronization of an all-to-all coupled $5-$cell network, into a splay state through the suppression of M-current.  Our work attempts to further the exploration of the desynchronization within the network, composed of intrinsically oscillating neurons, under different types of coupling architectures . We also explore the possible desynchronized states for each case, in the presence of M-current, associated with a low level of ACh.

The layout of this article is as follows. In Section \ref{Methodology}, we review reduction of general high-dimensional conductance-based model neuron to a one-dimensional phase model.  We also discuss how we use this phase model to study the cluster solutions of a biophysical network of conductance-based model neurons equipped with circulant coupling. In particular, we review the existence and stability conditions for networks equipped with an all-to-all globally homogeneous, symmetric distance-dependent and nearest-neighbours coupling. We introduce a single-compartment model of an excitatory hippocampal pyramidal neuron with an M-current, which is a reduction due to \parencite{olufsen_new_2003} of a detailed model developed by Traub and Miles \parencite{traub_simulation_1982}. We also perform a bifurcation analysis of this model to demonstrate the M-current induced switch in neuronal excitability, and propose model parameters that are appropriate for our study. In Section \ref{Results}, we present our results. We describe a phase model approximation of the RTM model neuron using parameters determined in the previous section, along with the stability analysis of various cluster solutions, under different strengths of the M-current, and the different coupling configurations of the corresponding network. Finally we summarize our work and discuss how our results may fit in the narrative of the cholinergic modulation of hippocampal activity in Section \ref{Discussion}. 
\section{Methodology}\label{Methodology}
In this section, we review the method to reduce a  weakly coupled network of neural oscillators to a phase model. We will then describe the solutions of the phase model, focusing on a particular class of solutions that describe the system in a phase-locked state.  Finally we introduce the specific neural network to which we apply the phase model analysis, along with the parameter values that are used in the numerical simulations.
\subsection{Phase Model Reduction}
\begin{figure}[h!]
     \centering
     \includegraphics[width=\linewidth]{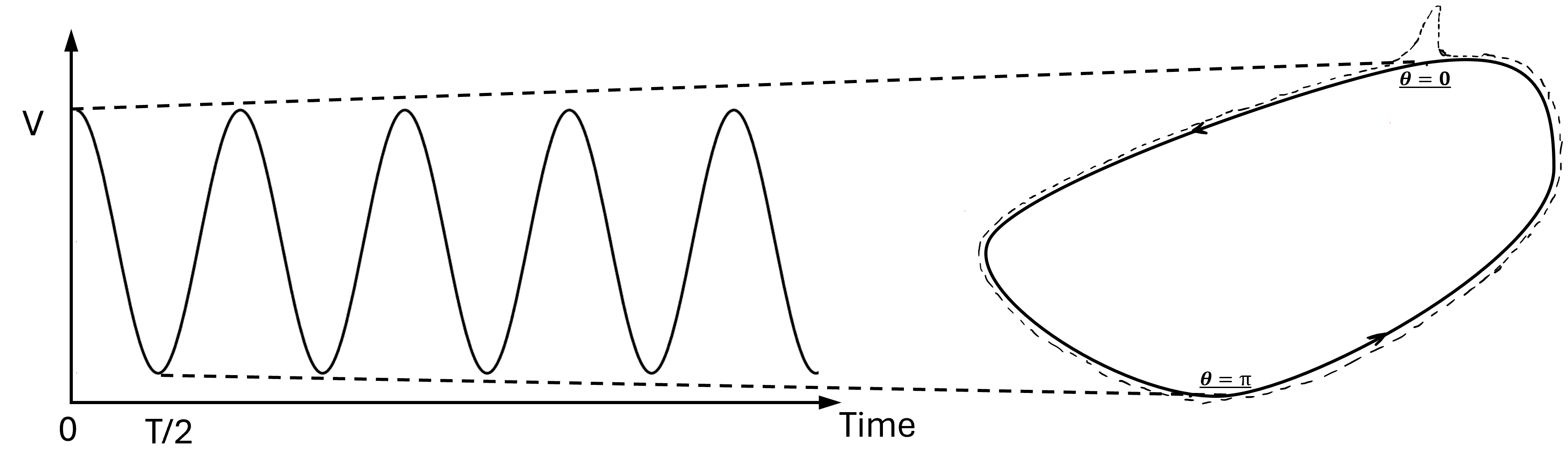}
     \caption{Representation of an oscillator exhibiting a stable period orbit in the uncoupled state. \textit{Left} Voltage trace showing the phase zero as the peak of the potential. \textit{Right} The stable orbit can be represented as a limit cycle in  phase space with $\theta=0$ and $\theta=\pi$ corresponding, respectively, to the spike times of the voltage trace and to times half a period, $T/2$, from the spike times.}
     \label{fig:phase}
 \end{figure}

    %Recall that 
    We represent the model for a single oscillator as follows:
\begin{equation}
    \frac{d{X}}{dt}=F(X), \label{eq1_singleoscl}
\end{equation}
where $X\in{\mathbb R}^n$. 
Assume that Eq.~\eqref{eq1_singleoscl} admits a unique exponentially asymptotically stable periodic solution $\hat{X}$ with period $T$. The adjoint system for Eq.~\eqref{eq1_singleoscl} about the solution, $\hat{X}$, is defined as follows:
\begin{equation}
    \frac{d{Z}}{dt}=-DF(\hat{X})^TZ.  \label{eq2_adj}
\end{equation}
We denote the unique periodic solution to this adjoint system as $Z=\hat{Z}(t)$ and assume that the following normalization condition is satisfied
\begin{equation}
    \int_{0}^T \hat{Z}(s)F(\hat{X}(s))\,ds =1.
\end{equation}
Now consider a network of identical oscillators given by %coupled through synaptic connections.
\begin{align}
     \frac{d{X_i}}{dt}=F({X_i})+\epsilon\sum^N_{j=1}w_{ij}G({X_i},{X_j}) \hspace{5mm} \label{eq4_networkoscl}
 \end{align}
 In Eq.~\eqref{eq4_networkoscl}, the coupling function denoted by $G: \mathbb{R}^N \times \mathbb{R}^N \longrightarrow \mathbb{R}^N$, describes the influence of oscillator $j$ on oscillator $i$. %X_i$ on $X_j$. 
 The strength of this influence is determined by the parameter $\varepsilon$. 
 $W=\{w_{ij}\}$  is the coupling matrix which describes the connectivity structure within the network, that is, information about which oscillators are coupled together. We follow the convention, which is appropriate for the context of neural oscillators, that $\epsilon$ is positive and the $w_{ij}$ are nonnegative. The sign of the coupling is determined by the function $G$. In the decoupled state ($\varepsilon =0$), each oscillator will satisfy Eq.~\eqref{eq1_singleoscl}. Consequently, when $\varepsilon << 1$ is sufficiently small, we can apply the theory of weakly coupled oscillators \parencite{hoppensteadt_weakly_1997} and obtain the following phase model approximation of Eq.~\eqref{eq4_networkoscl}
 \begin{equation}
     \frac{d\theta_i}{dt}=1 + \varepsilon \sum_{j=1}^Nw_{ij}H(\theta_i-\theta_j).\label{eq5_phasemodel}
 \end{equation}
In Eq.~\eqref{eq5_phasemodel}, $\theta_i$ is the phase of the oscillator represented by $X_i$, when the system in Eq.~\eqref{eq4_networkoscl} is exhibiting a stable periodic solution, $\hat{X}$. The relationship between $\theta_i$ and $X_i$,  when $\varepsilon=0$, is depicted in Fig~\ref{fig:phase}.
  $H$ is known as the interaction function 
 and can be expressed 
 as a convolution of the solution $Z(t)$ to the adjoint system Eq.~\eqref{eq2_adj} and the coupling function $G$ applied to the stable periodic solution $\hat{X}$ and this solution phase-shifted by an amount $\theta$:
 \begin{equation}
     H(\theta)=\frac{1}{T}\int_0^{T} Z(s)G(\hat{X}(s+\theta),\hat{X}(s))\,ds. \label{thmH}
 \end{equation}
 
 We assume that the network satisfies periodic boundary conditions. This corresponds to the neurons being arranged in a one dimensional ring. We further imposes symmetry on the network's coupling architecture. Specifically, we assume the coupling matrix $W$,  is circulant:
\begin{equation}
    W=\text{circ}\begin{pmatrix}
        w_0&w_1 & w_2 & \dots &w_{N-1}. \label{circ}
    \end{pmatrix}
\end{equation}
This can be written equivalently as
\begin{align}
    w_{ij}=w_{(j-i)\text{mod N}}.\label{sym}
\end{align}
In the corresponding biophysical neuron model, this means that the coupling between neurons does not depend on the particular neurons, only their relative positions in the network. Sometimes this is called distance-dependent coupling. %type of synaptic coupling structure is identical for each neuron, resulting in a rotational symmetry. 
Following the notation from \parencite{miller_clustering_2015}, if each neuron is coupled to $r$ of its nearest neighbours on either side, we say the network has a connectivity radius $r$, that is $w_k,w_{N-k}>0$ for all $k\leq r$. Additionally, we assume that there is no self-coupling ($w_0 =0$).

%Neglecting the higher order terms in $\varepsilon$, 
To facilitate our study of synchronization, we  introduce the following phase differences,
\begin{equation}
    \psi_i=\theta_{i+1}-\theta_i \hspace{5mm} i\mod{N}. \label{eq6_phasediff}
\end{equation}
The phase difference, $\psi_i$, describes relative phase %the relative spike times %of position in phase space 
of the $i+1^{st}$ oscillator with respect to that of ${i}^{th}$ oscillator. If this difference is $0$, then two oscillators are in-phase, that is, they are synchronized. The phase differences, Eq.~\eqref{eq6_phasediff}, must satisfy the periodic boundary conditions for Eq.~\eqref{eq5_phasemodel} since the model neurons are arranged in a one-dimensional ring. This gives the following constraint
\begin{equation}
    \sum^N_{i=1}\psi_i = 0 \text{ mod 2$\pi$} \label{sumconst}
\end{equation}
We may rewrite Eq.~\eqref{eq5_phasemodel} solely in terms of the phase differences, $\psi_i$, as follows
\begin{align}
    \frac{d\psi_i}{dt}&= \frac{d\theta_{i+1}}{dt}-\frac{d\theta_i}{dt}\nonumber\\
    % &=\varepsilon \left(\left(\sum^N_{j=1}w_{j-i-1(\text{mod N})} H(\theta_j-\theta_{i+1})\right)-\left(\sum^N_{j=1}w_{j-i(\text{mod N})} H(\theta_j-\theta_{i})\right)\right)\nonumber\\
    % &=\varepsilon \left(\left(\sum^{N-1}_{k=1}w_{k} H(\sum^{k-1}_{s=0} \phi_{s+i+1\mod{N}})\right)-\left(\sum^{N-1}_{k=1}w_{k} H(\sum^{k-1}_{s=0} \phi_{s+i\mod{N}}\right)\right)\nonumber\\
    &=\varepsilon \left(\sum^{N-1}_{k=1}w_k\left( H\left(\sum^{k-1}_{l=0}\psi_{l+i+1(\text{mod N})}\right)-H\left(\sum^{k-1}_{l=0}\psi_{l+i(\text{mod N})}\right)\right)\right)\label{eq6_phasedif2}
\end{align}
We call this the phase difference model.
\subsection{Existence and Stability of Cluster Solutions.}\label{sec:Clusterstab}
Equilibrium solutions of the phase difference model Eq.~\eqref{eq6_phasedif2} correspond to steady states in which the phase differences, $\psi_i$, remain constant. In Eq.~\eqref{eq5_phasemodel}, constant phase differences correspond to solutions $\theta_i=\Omega t+\theta_{i0}$. Thus the $\theta_i$ vary in time, but differ by fixed constants.  These solutions represent phase-locked states of the full system in Eq.~\eqref{eq4_networkoscl}.  We consider a particular class of phase-locked state in which oscillators of the network break up into subgroups of synchronized oscillators (i.e., with phase differences, Eq.~\eqref{eq6_phasediff}, equal to zero).
% \begin{figure}
%     \centering
%     \includegraphics[width=0.5\linewidth]{defense_phaselock.png}
%     \caption{Caption}
%     \label{fig:placeholder}
% \end{figure}
Necessary conditions for stability of any solution to the general model in Eq.~\eqref{eq5_phasemodel} have been studied in \parencite{galan_predicting_2006}. In this article, we focus on necessary and sufficient conditions for a specific type of solution, those that are model-independent, i.e., solutions that are independent of the interaction function $H$ and the weights, $w_{i}$,  in Eq.~\eqref{eq6_phasedif2}. 

Consider the constant vector given by
\begin{align} {\psi}_{eq} = \begin{bmatrix} \psi&\psi&\dots&\psi\end{bmatrix}, \label{eqiphase}\end{align} 
where the individual phase differences, $\psi_i$,  are identical, i.e, equal to $\psi$.  Clearly Eq.~\eqref{eqiphase} is a model-independent equilibrium solution of Eq.~\eqref{eq6_phasedif2}, since it satisfies these equations for any choices of the interaction function, $H$ and the coupling matrix, $W$.  Eq.~\eqref{eqiphase} corresponds in the full system Eq.~\eqref{eq5_phasemodel} to a periodic solution in which the neurons are phase-locked and the phase difference between any adjacent neurons is the same, $\psi$.  Applying the constraint equation in Eq.~\eqref{sumconst}, to Eq.~\eqref{eqiphase}, imposes restrictions on the choice of $\psi$ that lead to solutions, that is
\begin{equation}
    \sum^N_{i=1} \psi_i = N\psi = 0 \mod{2\pi} \implies \psi = 2\frac{k \pi}{N} \hspace{1cm} k \in \{0,1,..,N-1\}\label{constpsi}
\end{equation}
\begin{figure}
    \centering
    \includegraphics[width=\linewidth]{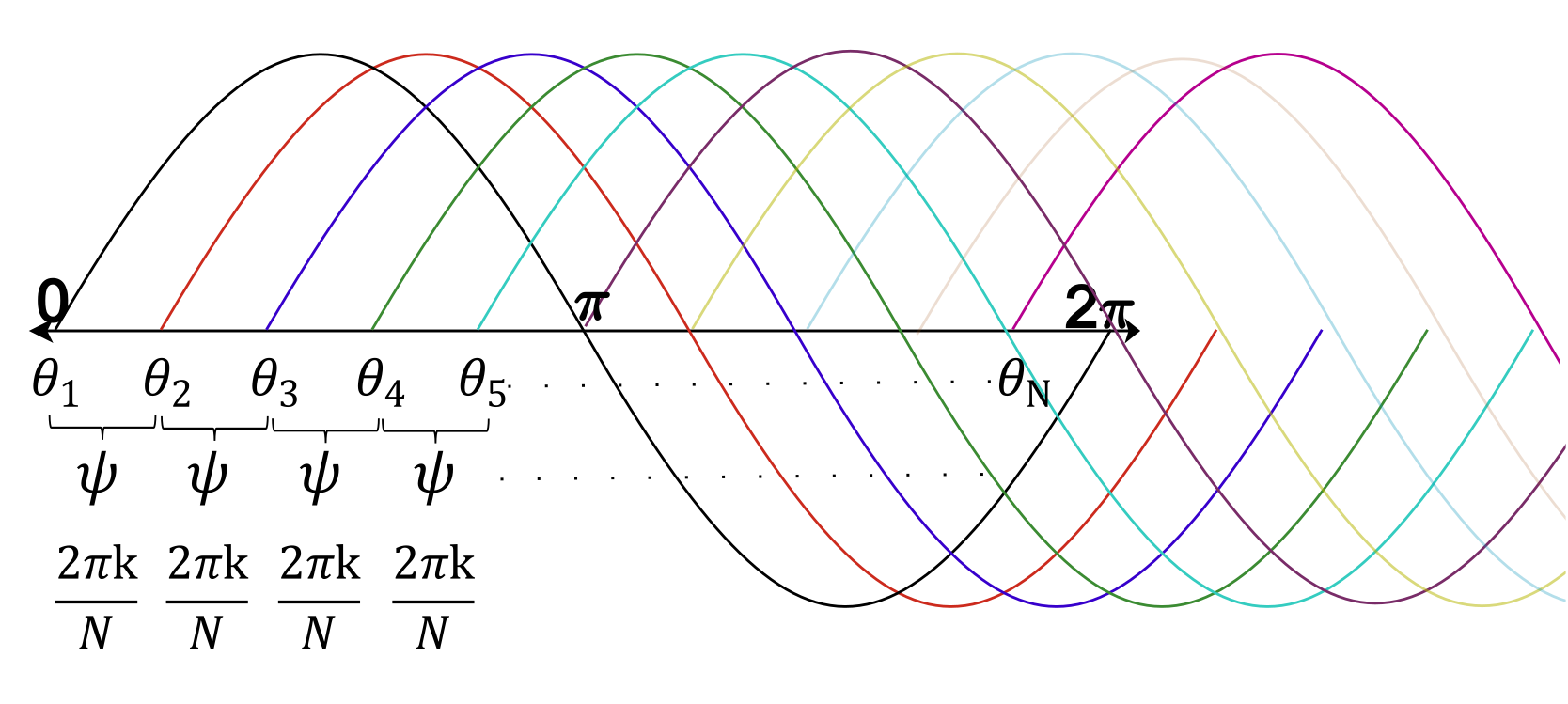}
    \caption{Voltage traces corresponding to an N-oscillator system exhibiting a symmetric $N-$cluster phase-locked or splay state solution where the adjacent phase differences are constant and identical.}
    \label{fig:phaselockpic}
\end{figure}
Based on  Eq.~\eqref{constpsi}, the phase difference equations, Eq.~\eqref{eq6_phasedif2}, for a network of size $N$ can admit $N$ solutions of the form Eq.~\eqref{eqiphase}. The values of $k$ and $N$ determine the relative phase differences between the different oscillators in corresponding phase-locked solution in the full system \eqref{eq4_networkoscl}.
For any value of $N$, the solution with $k=0$, i.e., $\psi=0$, corresponds in the full system to the state where all the oscillators are synchronized, which we also call the \textit{1-cluster} solution. 
For any $N$, values of $k$ such that $k$ and $N$ are relatively prime lead to solutions which correspond in the full system to each oscillator being in its own cluster, with phase difference $\psi=2k\pi/N$ between adjacent oscillators. This corresponds to the \textit{N-cluster} solution, also called the splay state.  This solution is illustrated in Fig.~\ref{fig:phaselockpic}. For $k,N$ that are not relatively prime, let $q={\rm gcd}(k,N)$. Then $N=qn$ and $k=qm$ for some integers $m,n$ and the equilibrium solution \eqref{constpsi} corresponds in the full system to an \textit{n-cluster} solution. Specifically, the oscillators split into $n$ groups/clusters, each containing $q$ oscillators. Within a cluster the phase difference between oscillators is $0$, while the phase difference between adjacent neurons in different clusters is  $2k\pi/N=2m\pi/n$.  So if $n$ divides $N$ then a network with $N$ neurons has \textit{n-cluster} solutions with phase differences $\psi=2m\pi/n$ where $m<n$ and ${\rm gcd}(m,n)=1$. Further details on the derivation of these results can be found in \parencite{miller_clustering_2015}.
Note that each of the solutions described above is characterized by clusters containing an identical number of oscillators. This particular type of cluster solution is called a \textit{symmetric cluster solution}. %In general, Eq.~\eqref{eq6_phasedif2} \textit{will always admit the in-phase or synchronized solution}. If $n$ divides $N$, then Eq.~\eqref{eq6_phasediff} admits an $n-$cluster solution in which the phase difference between adjacent neurons is given by  $\psi=\frac{2\pi m}{n}$ for $m<n$. 
%Further details of the derivation of the different types of symmetric clusters that may emerge can be found in \parencite{miller_clustering_2015}.

To investigate the stability of the symmetric cluster solutions described above, we linearize the system Eq.~\eqref{eq6_phasedif2} about the equilibrium solution Eq.~\eqref{eqiphase}. The eigenvalues of the linearization can be calculated exactly as shown in \parencite{campbell_phase_2018}, where they proved the following result.
The equilibrium point $\psi =\frac{2\pi k}{N}$ $k \in \{0,1, \dots,N-1\}$, and the corresponding cluster solution for the phase difference model Eq.~\eqref{eq6_phasedif2} is asymptotically stable if the $\mu_j$ defined below is positive for all $1<j< \lfloor\frac{N}{2}\rfloor$ 
\begin{align} 
 \mu_j=\sum^{N-1}_{l=1}w_lH'(l\psi)\left(1-\cos\left(\frac{2\pi lj}{N}\right)\right)\label{bi-dire_eig}\end{align}
and is unstable if at least one of the $\mu_j$ is negative. 

Depending on the choice of the coupling radius $r$ and the values of $w_k$, we consider how the expression for $\mu_j$ simplifies for specific configurations of our system, two common simple connectivity structures and a more general one. 

First, however, we consider the synchronized ($1$-cluster) solution. Since $\psi=0$ in this case, Eq.~\eqref{bi-dire_eig} simplifies to 
\begin{align} 
 \mu_j=H'(0)\sum^{N-1}_{l=1}w_l\left(1-\cos\left(\frac{2\pi l}{N}\right)\right).\label{synch}\end{align}
Since $w_l$ and $\left(1-\cos\left(\frac{2\pi l}{N}\right)\right)$ are nonnegative, we have the following result.

\textit{For any network structure, the synchronous solution, i.e., $1-$cluster solution, is asymptotically stable if $H'(0)>0$ and unstable if $H'(0)<0.$}
 
\subsubsection{Networks with \textit{nearest neighbour} coupling}\label{sec:NNstab}
This simple choice of coupling occurs when each neuron is only connected to two other neurons, those directly adjacent. This configuration is obtained by setting the coupling radius, $r$, to 1, giving the coupling matrix 
$W =w_1\cdot circ\begin{pmatrix}
            0&1&0&\cdots &0 & 1
        \end{pmatrix}$.   For  $j\in\{1,2,\dots,\lfloor\frac{N}{2}\rfloor\}$, $\mu_j$ in Eq.~\eqref{bi-dire_eig} reduces to the following
\begin{align}\label{nneig}
   \hspace{-.2in} \ \mu_j=w_1\left(H'(\psi)+H'(-\psi)\right)\left(1-\cos\left(\frac{2\pi j}{N}\right)\right)=2w_1H'_{odd}(\psi)\left(1-\cos\left(\frac{2\pi j}{N}\right)\right)
\end{align}
In Eq.~\eqref{nneig}, $w_1$ and $\left(1-\cos\left(\frac{2\pi j}{N}\right)\right)$ are nonnegative. In order to guarantee stability, we only need to consider the expression $H'(\psi) + H'(-\psi) $ which is independent of $j$. Denoting the odd part of $H$ by $H_{odd}$ yields the following result.

% \begin{corollary}\label{NNminCor}
\textit{    A symmetric cluster solution associated with the equilibrium point $\mathbf{\psi}$ for a network equipped with nearest neighbour coupling is asymptotically stable if ${H'_{odd}(\psi) > 0}$ and unstable if  ${H'_{odd}(\psi) < 0}$.}%and only if ${H'_{odd}(\psi) > 0}$.}
% \end{corollary}
\subsubsection{Networks with all-to-all \textit{global homogeneous }}
This choice of coupling is a special case of all-all coupling, in which all the weights $w_j$ in Eq.~\eqref{bi-dire_eig} are the same, i.e., the coupling matrix is 
$W =w_1\cdot circ\begin{pmatrix}
            0&1&1&\cdots &1 & 1
        \end{pmatrix}$. 
In this case, the symmetric cluster solution associated with the equilibrium point $\psi$ is asymptotically stable if and only if
\begin{equation} \mu_j=w_1\sum^{N-1}_{l=1}  H'(l\psi)\left(1-\cos{\frac{2\pi lj}{N}}\right) > 0 \hspace{1cm} \forall j \in \left\{1,..,\big\lfloor\frac{N}{2}\big\rfloor\right\}.\label{mujall}\end{equation}
This set of inequalities can be rewritten in a form that %modified such that the conditions outlined in Eq.~\eqref{mujall} 
is independent of the size of the network $N$ and the equilibrium point value $\psi$, instead depending on the {\em number of clusters} $n$: %:  
\begin{align}\label{mujall1}
     &\mu_j=\sum^{n-1}_{l=1}  \left(H'\left(\frac{2\pi l }{n}\right)\right)\left(1-\cos{\frac{2\pi lj}{n}}\right), \hspace{1cm} 1\leq j \leq \big\lfloor\frac{n}{2}\big\rfloor \nonumber\\\\
      &\hat\mu_0 =  \sum^{n-1}_{l=0}  \left(H'\left(\frac{2\pi l }{n}\right)\right). \nonumber
\end{align}
%Essentially, we can rewrite $\mu_j$ in Eq.~\eqref{mujall} to generalize the stability analysis of a symmetric $n-$cluster solution where $n<N$.

Eq.~\eqref{mujall1} recovers the results from \parencite{okuda_variety_1993}. It implies that the stability of a $n-$cluster solution is independent of the size of the network. Further, since all the coupling weights are identical, networks with global homogeneous coupling are unaffected by any rearrangement of indices \parencite{ashwin_dynamics_1992,wang_phase_2015}.
Thus {\em any} symmetric $n-$cluster solution of \eqref{eq4_networkoscl}, is equivalent to the one defined by \eqref{eqiphase}-\eqref{constpsi} by rearrangement of the indices.
It follows that, Eq.~\eqref{mujall1} determines the stability for any symmetric $n-$cluster solution in a network equipped with a globally homogeneous coupling. 
\subsubsection{Networks with all-to-all \textit{distance-dependent} coupling}
In real neural networks, the coupling strengths are likely to be heterogeneous. Thus, it is worthwhile to study configurations in which all the weights $w_i$ in Eq.~\eqref{bi-dire_eig} are different, but not random. To do thi, we introduce a \textit{selective heterogeneity} in the synaptic weights. In particular, we consider a network where the relative coupling strengths for a neuron decrease with distance from the neuron. This does not necessarily reflect a physical distance in the corresponding network, but rather a hierarchy in the strengths of the coupling between neurons in the network. We call this type of coupling distance-dependent \parencite{campbell_phase_2018}.  The corresponding coupling matrix $W$ will satisfy Eq.~\eqref{bi-dire_eig} with  $w_{i}= w_1\cdot p^{i-1}$ for some non zero $p < 1$:
\begin{equation}
    W =w_1\cdot circ\begin{pmatrix}
            0&1&p&p^2&\cdots &p^2&p & 1
        \end{pmatrix} \label{eq:DDcoupling}
\end{equation}
       As such, Eq.~\eqref{bi-dire_eig} reduces to the following \parencite{campbell_phase_2018} :  
\begin{equation} \mu_j=w_1\cdot\sum^{N-1}_{l=1} p^{l-1} (H_{odd}'(l\psi)+H_{odd}'(-l\psi))\left(1-\cos{\frac{2\pi lj}{N}}\right) > 0, %\hspace{0.5cm} \forall j \in \{1,..,\big\lfloor\frac{N}{2}\big\rfloor\},
\label{mujDDodd}\end{equation}
for $ j \in \{1,..,\big\lfloor\frac{N}{2}\big\rfloor\}$ with $N$ odd and 
\begin{align}\label{mujDDeve}
&\mu_j=w_1\cdot\left( p^{\frac{N-3}{2}}H\left(m\psi\right)\left(1-(-1)^j\right) +2\sum^{m-1}_{l=1} p^{l-1} (H_{odd}'(l\psi)+H_{odd}'(-l\psi))\left(1-\cos{\frac{2\pi lj}{N}}\right) > 0\right) \nonumber\\
%& \hspace{12cm}\forall j \in \{1,..,m\},
\end{align}
for  $j \in \{1,..,m\}$ with $N=2m$, even.

Unlike the nearest-neighbours and global homogeneous cases, the stability of an equilibrium point $\psi$ associated with a cluster solution for the system in Eq. ~\eqref{eq6_phasedif2}, equipped with distance-dependent coupling, \textit{will depend on the size of the network}. 

As we can see, for arbitrary values of $w_j$, we cannot derive a minimal stability condition like in the case of nearest neighbours. Thus, to determine stability we will be required to compute the values of  all the $\mu_j$ in Eq.~\eqref{mujDDeve} or Eq.~\eqref{mujDDodd}, associated with a cluster solution. %This leads us to consider a special case of all-all coupling where minimal stability conditions can be computed.

\subsection{Neuron Model}\label{MOdel}
To model the pyramidal neuron, we use a  single compartment, conductance-based model \parencite{olufsen_new_2003}. This model is simplification of a multi-compartment model for a pyramidal excitatory cell in rat hippocampus due to Traub and Miles \parencite{traub_model_1991}, with the addition of a muscarinic current (M-current). We refer to this as the Reduced Traub-Miles (RTM) model.
The parameters are taken from \parencite{ermentrout_fine_1998, kopell_gamma_2000,crook_role_1997,olufsen_new_2003}. The equations for the model are
\begin{align}
    C\frac{dV}{dt}&= I_{app}-g_L\cdot(V-V_L)-{g}_k\cdot n^4\cdot (V-V_K)-g_{Na}\cdot m^3\cdot h\cdot (V-V_{Na}) \label{K&O}\\
    &\hspace{.15in}-g_M\cdot w\cdot(V-V_K), \\
    \frac{dm}{dt}&=\alpha_m(V)(1-m)+\beta_m(V)m\\
    \frac{dn}{dt}&=\alpha_n(V)(1-n)+\beta_n(V)n\\
    \frac{dh}{dt}&=\alpha_h(V)(1-h)+\beta_h(V)h\\
    \frac{dw}{dt}&=\frac{w_\infty(V)-w}{\tau_w(V)}\label{w_eq}
\end{align}
where $V$ is the membrane potential in mV and $t$ is the time in ms. The dimensionless gating variables $h$ and $m$ represent the inactivation and activation of the sodium current, respectively. The dimensionless gating variables $n$ and $w$ represent the activation of the delayed rectifier potassium and muscarinic currents, respectively. The reaction rates associated with activation of the sodium and potassium channels are  $\alpha_m(V)= {0.32(54+V)}/({1-\exp{\left(-({54+V})/{4}\right)}}) $,$\beta_m(V)={0.28(V+27)}/{(\exp\left({({V+27})/{5}}\right)-1)}$, $\alpha_n={0.032(V+52)}/{1-\exp\left({V+52}/{5}
    \right)}$,  $\beta_n=0.5\exp{\left({-(57+V)}/{40}\right)}$. The rates are voltage dependent with units ms$^{-1}$. Similarly, the reaction rates associated with inactivation of the sodium channel are $\alpha_h(V)= 0.128\exp{\left(({-(V+50)})/{18}\right)}$ and $\beta_h(V)={4}/{(1+\exp{\left((-{V+27})/{5}\right)})}$. The steady state function and time constant associated with the activation of the muscarinic channel are respectively given by $w_{\infty}(V)={1}/({1+\exp{\left(({-(V+35)})/{10}\right)}})$ and $\tau_{w}(V)={400}/({3.3\exp{\left(({V+35})/{20}\right)}+\exp{\left(-({V+35})/{20}\right)}})$. 
    The maximal conductances for the sodium, potassium and leak currents are $g_{Na}=100$ {mS/cm}$^2$, $g_{K}=80$ {mS/cm}$^2$ and $g_{L}=0.1$ {mS/cm}$^2$, respectively. The corresponding reversal potentials of the currents are $V_{Na}=50$ mV, $V_{K}=-100$ mV, $V_L=-67$  mV, while the capacitance is $C=1 $ $\mu F/cm^2$.
    
\subsubsection{Modelling the effect of Acetylcholine (ACh)}\label{sec:AChmodel}
In our study, we vary the maximal conductance of the M-current, $g_M$, to represent the effect of acetylcholine on the pyramidal neurons modelled with Eqs.~\eqref{K&O}-\eqref{w_eq}. %In fact 
Cholinergic  suppression of the M-current has been shown to switch pyramidal neurons from exhibiting class II to class I excitability both computationally \parencite{fink_cellularly-driven_2011} via reduction of $g_M$ and experimentally \parencite{stiefel_cholinergic_2008}. Analysis of conductance-based models attributed this switch in neuronal excitability to the presence of a codimension-2 Bogdanov-Takens (BT) bifurcation, which arises through the variation of the parameters, $I_{app}$ and $g_M$ \parencite{ermentrout_mathematical_2010,al-darabsah_m-current_2021}. 
 
\begin{figure}
    \centering
    \includegraphics[width=\linewidth]{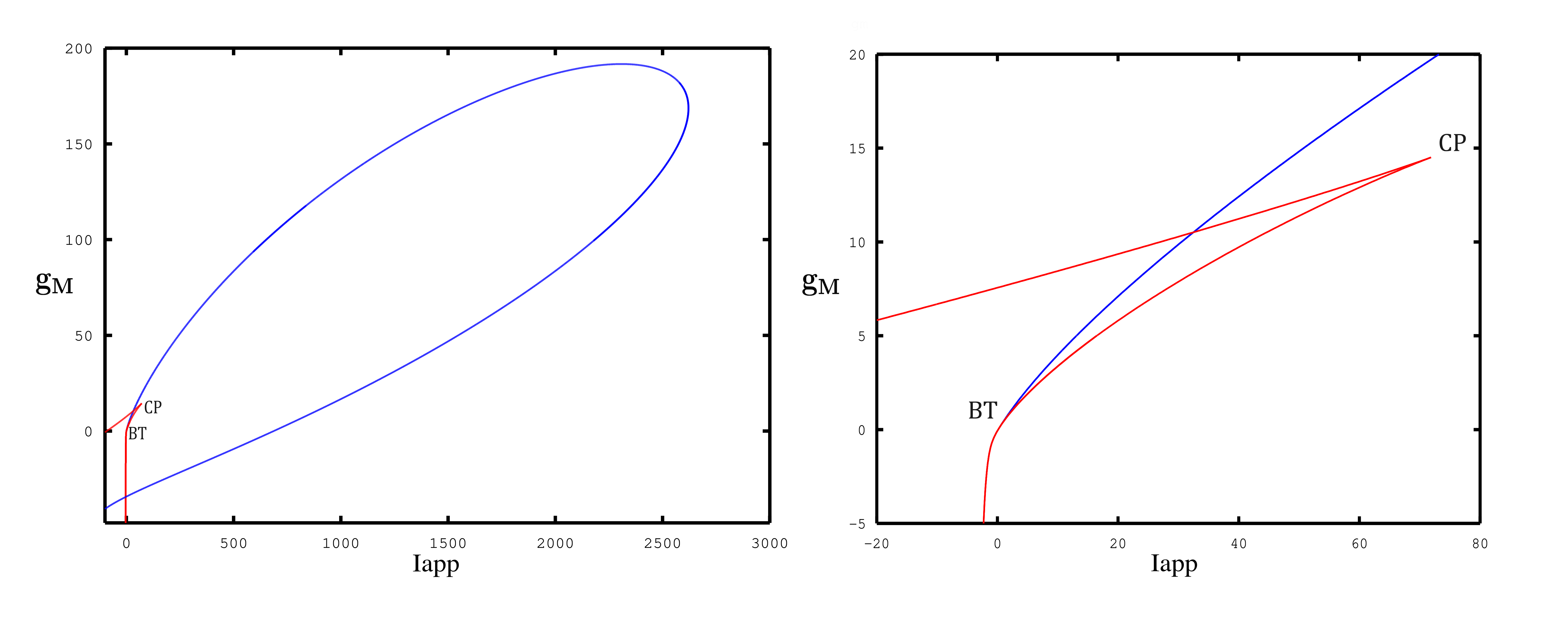}
    \caption{Bifurcation curves in (${I_{app}, g_M}$) parameter space. Red curves correspond to fold bifurcation of equilibria, blue to Hopf bifurcation. (A). Full two-parameter bifurcation diagram (B). Zoomed-in view near the red curves, highlighting the region relevant to our analysis. As $g_M$ increases, the two fold curves merge at a cusp point (labelled CP). Additionally, we see a curve of Hopf points emerging from the BT point. }
    \label{fig:2parbif}
\end{figure}

\begin{figure}
    \centering
    \includegraphics[width=1\linewidth]{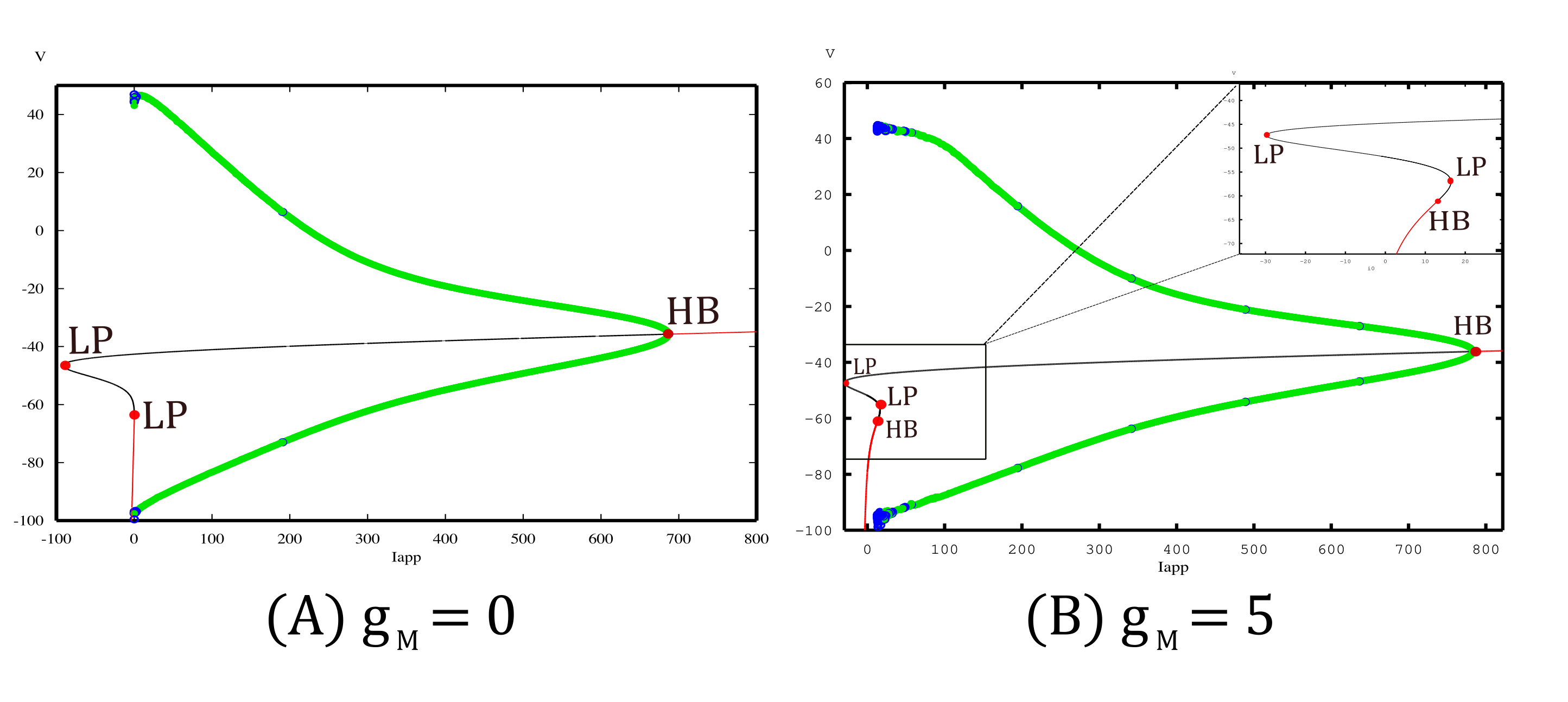}
    \caption{\footnotesize One parameter bifurcation diagrams as a function of current ($I_{app}$) for the RTM model. The red (black) curves indicate stable (unstable) equilibrium points. The green (blue) curves show the maximum and minimum voltages along the stable (unstable) periodic orbits. \textbf{(A)} Below the BT bifurcation point in Figure~\ref{fig:2parbif} ($g_M=0$). The oscillations are born via a saddle-node on an invariant circle as $I_{app}$ is increased. Thus the model exhibits class I excitability.  \textbf{(B)} Above the BT bifurcation point ($g_M=5$). The oscillations are born via a subcritical Hopf bifurcation (labelled HB). Although the limit points still exist, the branch of periodic solutions are emerging from the Hopf point rather than the limit point. Thus the model exhibits class II excitability. Numerical limitations prevented us from demonstrating the connection between the blue curves.}

    \label{fig:bifIV}
\end{figure}

 In Figure~\ref{fig:2parbif} and Figure~\ref{fig:bifIV}, we reproduce the bifurcation diagrams of \parencite{al-darabsah_m-current_2021}, for the RTM model, Eqs.~\eqref{K&O}-\eqref{w_eq}.
 %in the $(I_{app},g_M)$ and $(I_{app},V)$ parameter space, respectively. 
Figure ~\ref{fig:2parbif}, shows two-parameter bifurcation curves in terms of the applied current, $I_{app}$, and the maximal conductance of the M-current, $g_M$. The transition from class I to class II occurs around the Bogdanov Takens point (labelled BT)  at $g_M \simeq 0.03$. Based on this bifurcation analysis, we choose representative values of $g_M$ where the system exhibits class I and II behaviours to be 0 and 5, respectively. %These curves additionally provide insight into suitable values of the externally applied current  ($I_{app}$) that allow the RTM model neuron to exhibit stable limit cycle oscillations for a given maximum value of $g_M$, which is a pre-requisite to obtaining a phase model reduction. 
Figure~\ref{fig:bifIV} shows one parameter bifurcation diagrams in terms of $I_{app}$, demonstrating a change in the bifurcation structure as the M-current is increased. In particular, we see that at $g_M=0$, when there is no M-current, the stable oscillations emerge as the stable equilibrium is destroyed in a saddle-node bifurcation. This corresponds to a SNIC bifurcation indicating class I excitability. This switches to class II excitability when $g_M$ is increased to $5$. In this case, we see oscillations emerge through a subcritical Hopf Bifurcation as a stable equilibrium becomes unstable. Figure \ref{fig:bifIV} also indicates that an applied current, $I_{app}$, value of $60mA$, is enough to initiate repetitive spiking behaviour in the model neuron for both values of $g_M$. In summary, we set $g_M=0$ to represent a neuron under the effect of a 'high' concentration of ACh. A neuron under the effect of a 'low' concentration of ACh is represented by increasing the parameter $g_M$ to $5$.

\subsubsection{Network Model}
To investigate this effect at a population level, we consider a network of identical neurons coupled with excitatory synapses. The synapses are represented using a standard first order kinetic model \parencite{destexhe_synthesis_1994}, giving the network model

\begin{align} 
    \frac{dV_i}{dt}&=I_{app}-g_L(V_i-V_L)-g_M w_i (V_i-V_K)-g_{Na}m_i^3h_i(V_i-V_{Na})\nonumber \\&-g_Kn_i^4(V_i-V_K) - \frac{g_{syn}}{N}\sum^N_{j=1}W_{ij} s_j(V_i-V_{syn})\nonumber \\\nonumber\\
    \frac{dX_i}{dt}&=\alpha_{X_i}(1-X_{i})+\beta_{X_i}X_{i} \hspace{2.5cm}X = \{m,h,n\} \label{network}\\ \nonumber \\ 
    \frac{dw_i}{dt}&=\frac{w_\infty(V_i)-w_i}{\tau_w(V_i)}\nonumber\\
    \frac{ds_i}{dt}&=5\left(1+\tanh{\left(\frac{V_i}{4}\right)}\right)(1-s_i)-\frac{s_i}{2} \hspace{1cm}i=\{1,2,..,N\}\nonumber
\end{align}

In Eq.~\eqref{network}, $g_{syn}$ is normalized by $N$ to reflect the individual contributions from the other cells.
As described in \parencite{kopell_gamma_2000,olufsen_new_2003}, we will assume the synapses are AMPA-mediated and accordingly, $V_{syn}$ is set to $0$ mV \parencite{ermentrout_mathematical_2010}. The other parameters of the synapse model are taken from \parencite{olufsen_new_2003}.

\section{Results}\label{Results}
\subsection{{Approximation of the Interaction Function}}
Except for a few special cases, it is generally not possible to analytically derive the interaction function $H$ \parencite{ermentrout_mathematical_2010}. We thus follow a standard numerical procedure for calculating this function. Using XPPAUT \parencite{ermentrout_simulating_2002}, we can numerically solve for the adjoint system, Eq.~\eqref{eq2_adj},  of a neuron exhibiting limit cycles and consequently their interaction function, Eq.~\eqref{thmH}. The interaction function $H$ can then be approximated by a truncated Fourier series that is readily available from XPPAUT: 
\begin{align}
    H(\phi)\approx a + a_1\cos{\phi} +b_1\sin{\phi} + a_2\cos{2\phi} +b_2\sin{2\phi}+  \sum^N_{i=3} ( a_n\cos{n\phi} +b_n\sin{n\phi}) \label{taylorex}
\end{align}
The interaction function $H$ was calculated in this manner for four distinct values of the maximal M-current conductance, $g_M$, each corresponding to a different strength of the M-current. As discussed in section~\ref{sec:AChmodel}, in this paper we focus on two specific values of $g_M$:
\begin{equation}
    g_M=\{0,5\}, \label{gmvalues}
\end{equation}
corresponding to high and low ACh, respectively. 

\begin{figure}
    \centering
    \includegraphics[width=\linewidth]{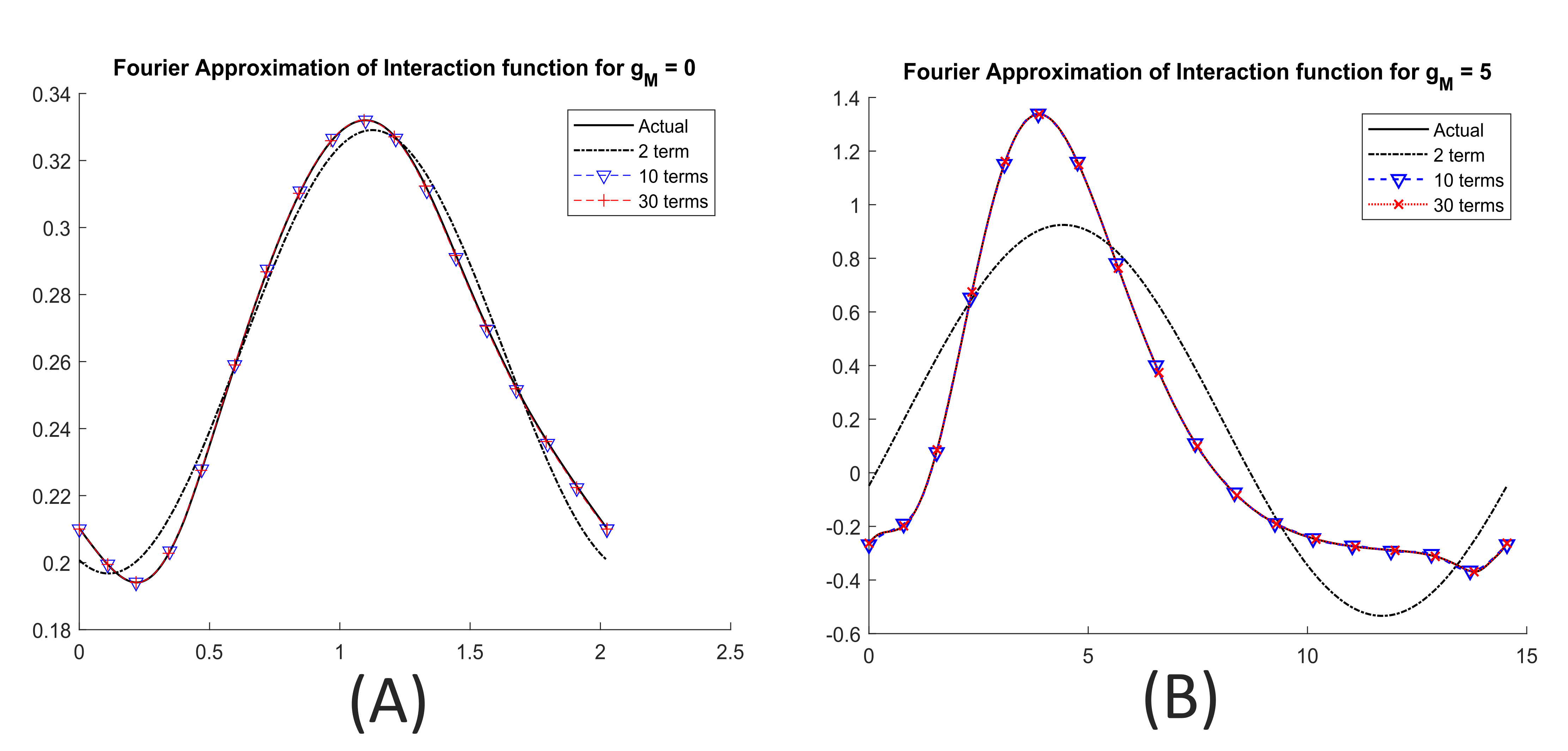}
    \caption{Interaction function for the reduced Traub-Miles model (N set to 2 in Eq.~\eqref{network}) evaluated at four distinct values of the M-current maximal conductance, $g_M$. We compare the different approximations in each case using $2$, $10$ and $30$ terms.  \textbf{(A)} $g_M=0$ (high ACh) \textbf{(B)}  $g_M=5$ (low ACh)}
    \label{fig:HAPPROX}
    \end{figure}

The numerical approximation of the interaction function corresponding to the two values of $g_M$ in Eq.~\eqref{gmvalues}, is summarized in Figure \ref{fig:HAPPROX}. Approximations using $2,10$ and $30$ terms of the Fourier series are compared. Adding further terms did not significantly improve the approximation, thus the $30$-term approximation was chosen to minimize the computational burden while maintaining accuracy in the evaluation of the interaction function, $H$. 
Note that the shape of the interaction function is drastically different for the two different values of $g_M$. This is a reflection of the different dynamics of the RTM model at these values discussed in section~\ref{eq1_singleoscl}.
\subsection{Phase model predictions for the excitatory  RTM network}
In this section we use the phase model to predict the stability of various cluster solutions of the model network, for $g_M$ corresponding to high and low ACh, in three network coupling configurations: global homogeneous, distance dependent and nearest neighbour. To do this we use the approximation of the interaction function from the previous section to calculate the quantities $\mu_j$ given in section~\ref{sec:Clusterstab}.   

\subsubsection{Networks with all-to-all \textit{globally homogeneous} coupling} 
Recall that in this situation, all of the synaptic weights, $w_i$, are equal to $w_1$ (with $w_0=0$), and that the value of $w_1$ does not influence the stability. Hence we take $w_1=1$.
Table~\ref{tab:psi stable Glob Hom}  shows the results when the stability criteria given in Eq.~\eqref{mujall1} are used to determine the stable symmetric $n-$cluster solutions for a network of arbitrary size. 
Columns $2-4$ in Table \ref{tab:psi stable Glob Hom} correspond to the network at a particular strength of the M-current (indicated by $g_M$) and each row summarizes the stability trend for a specific $n-$cluster.  For a given value of $n$, we compute Eq.~\eqref{mujall}  for each $g_M$, and if it is positive, we mark the corresponding $n-$cluster as a predicted stable solution. %for some appropriate network at particular strength of the  M-current.
Note that these predictions are independent of the number of neurons, $N$, in the network. For any $N$ that is divisible by $n$, the corresponding network with $N$ neurons admits a symmetric $n-$cluster solution, with predicted stability as given in Table~\ref{tab:psi stable Glob Hom}.

\begin{table}[h!]
    \centering
     \begin{tabular}{ccc}
     \hline
         {Cluster Size}&\multicolumn{2}{c}{Value of $g_M$}\\
        % \cline{2-3} 
         ${n}$&${ 0 }$&  ${ 5 }$\\\hline\hline
      $15$&{Unstable}&  {Unstable}\\
     $14$&{Unstable}&  {Unstable}\\
     $13$&{Unstable}&  {Unstable}\\
           $12$&{Unstable}& {Unstable}\\
     $11$&{Unstable}&  {Unstable}\\
             $10$&{Unstable}&{Unstable}\\
     $9$&{Unstable}&  {Unstable}\\
              $8$ &{Unstable}& {Unstable}\\
     $7$&{{Stable}$^*$}&  {Unstable}\\
             $6$ &{{Stable}$^*$}&{Unstable}\\
     $5$ &{{Stable}$^*$}&  {Unstable}\\
     $4$ &{{Stable}$^*$}&  {Unstable}\\
     $3$ &{Unstable}&  {Unstable}\\
     $2$&{Unstable}&   {Unstable$^*$}\\
              1&{Unstable$^*$}&   {{Stable}$^*$}\\
              \hline
    \end{tabular}
 \vspace{1mm}
 \caption{Phase model predictions for $g_M =0$ (high ACh) and $g_M=5$ (low ACh) in RTM networks with \textit{global homogeneous coupling}. The solutions that were verified through numerical simulation of the full network model are marked with a $^*$.}
 \label{tab:psi stable Glob Hom}
\end{table}

The phase model predicts that the in-phase solution ($n=1, \psi =0$) for a network of any size is unstable when the M-current is zero and then becomes stable for a M-current with a conductance parameter $g_M=5$. This trend in the stability is identical to that for a pair of neurons observed in \parencite{al-darabsah_m-current_2021}, due to the fact that the stability criteria for the in-phase ($n=1,\psi=0$) is independent of the size of the network $N$ for {any} coupling architecture. See note just after Eq.~\eqref{synch}. We expect this stability trend to be consistent across all networks and coupling architectures satisfying periodic boundary conditions. The in-phase solution is also the only symmetric solution predicted to be stable when the M-current conductance is $g_M=5$.  When $g_M=0$, the phase model predicts that there are multiple stable $n-$cluster solutions ranging from $4$ to $7$ clusters. 

\subsubsection{Networks with all-to-all \textit{distance-dependent} coupling}
Now we turn to all-to-all coupling with some structural heterogeneity, in particular distance-dependent coupling. Recall that the coupling weights in this situation are given by Eq.~\eqref{eq:DDcoupling}. We set the parameter $p=0.5$, which corresponds to relative coupling strength ($w_{i+1}/w_{i}=0.5$), and $w_1=1$.  As discussed in section~\ref{sec:Clusterstab}, for a network of size $N$ the admissible symmetric cluster solutions are ones where adjacent cells are spiking with a phase difference $\psi=\frac{2\pi k}{N}$ for some $k \in \{1,2,..,N\}$. Using either Eq.~\eqref{mujDDodd} or Eq.~\eqref{mujDDeve} as appropriate, the stability was determined for various cluster solutions in networks with $N=4,5,6,8,10$, $12,20$ and $24$. 
The resulting stability predictions are summarized in Table~\ref{tab:AA DD 0.5 psi stable}. Each row in Table~\ref{tab:AA DD 0.5 psi stable} shows the  stability for a given ($N,\psi$) pair for the two choices of M-current conductance strength, $g_M$. As expected from Eq.~ \eqref{synch}, the stability trend for the in-phase is identical to that of a pair of weakly coupled neurons. When $g_M=5$, up until $N=10$, the phase model predictions are identical to the case of the global homogeneous coupling with the in-phase state being the only stable solution. Beyond $N=10$, the phase model predicts that at least one splay solution is stable in addition to the synchronized state. When $g_M=0$, the phase model predicts that there are multiple stable $n-$cluster solutions. 

In contrast with the global homogeneous case, for a given $g_M$, a particular $n-$cluster solution can differ in stability for the same size network ($N$) but with different values of $\psi$. For example, consider $N=8$, when there is no M-current ($g_M=0$), the phase model predicts that an $8-$cluster solution (splay state) with $\psi=\frac{\pi}{4}$ is unstable but stable if $\psi=3\frac{\pi}{4}$. An  $n-$cluster solution  associated with the same $\psi$ can also differ in stability across networks with different sizes, $N$.  As an example, consider $N=12$ in Table \ref{tab:AA DD 0.5 psi stable}, when $g_M=5$, the phase model predicts that a $12-$cluster with $\psi=\frac{\pi}{6}$ is stable. The same solution is predicted to be unstable for $N=24$  at $g_M=5$.  

\begin{table}[h!]
\centering 
  \begin{tabular}{ccccc}\hline
     Network Size&Phase Difference&\multicolumn{2}{c}{Value of ${g_M}$}&Cluster Size\\
     
     $N$& $\mathbf{\psi}$& ${ 0 }$&  ${ 5 }$&  ${n}$\\\hline \hline
\renewcommand{\arraystretch}{0.9}
 \multirow{3}*{$4$}& $0$& {Unstable}$^*$&  {Stable}$^*$&$1$\\
 & $\frac{\pi}{2},3\frac{\pi}{2}$& {Stable}$^*$&  {Unstable}&$4$\\
 & $\pi$& {Stable}$^*$&  {Unstable}&$2$\\ \hline 
     \multirow{3}*{$5$}&$0$&  {Unstable}$^*$&{Stable}$^*$&$1$\\
          &$2\frac{\pi}{5},8\frac{\pi}{5}$& {Stable}&{Unstable}& $5$\\
          &$4\frac{\pi}{5},6\frac{\pi}{5}$& {Stable}$^*$& {Unstable}& $5$\\\hline
 \multirow{5}*{$6$}& $0$& {Unstable}$^*$&  {Stable}$^*$&$1$\\
 & $\frac{\pi}{3},5\frac{\pi}{3}$& {Stable}&  {Unstable}&$6$ \\
 & $2\frac{\pi}{3},4\frac{\pi}{3}$& {Stable}$^*$&  {Unstable}&$3$ \\
 & $\pi$& {Stable}&  {Unstable}&$2$\\
    \hline
\multirow{6}*{$8$}&$0$&  {Unstable}$^*$&{Stable}$^*$&1\\
      &$\frac{\pi}{4},7\frac{\pi}{4}$&   {Unstable}& {Unstable}& $8$\\
          &$\frac{\pi}{2},3\frac{\pi}{2}$& {Stable}$^*$&{Unstable}& $4$ \\
 & $3\frac{\pi}{4},5\frac{\pi}{4}$& {Stable}$^*$&  {Unstable}&$8$\\
          &$\pi$&  {Stable}$^*$&  {Unstable}&$2$\\
          \hline
\multirow{7}*{12}&$0$&  {Unstable}$^*$&{Stable}$^*$&1\\
      &$\frac{\pi}{6},11\frac{\pi}{6}$&   {Unstable}& {Stable}$^*$& $12$\\
          &$\frac{\pi}{3},5\frac{\pi}{3}$& {Stable}&{Unstable}& $6$ \\
          &$\frac{\pi}{2},3\frac{\pi}{2}$& {Stable}$^*$&{Unstable}& $4$ \\
          &$5\frac{\pi}{6},7\frac{\pi}{6}$& {Stable}& {Unstable}& $12$ \\
          &$2\frac{\pi}{3},4\frac{\pi}{3}$& {Stable}$^*$& {Unstable}& $3$ \\
          &$\pi$&  {Stable}$^*$&  {Unstable}&$2$\\\hline
 \multirow{6}*{$16$}& $0$& {Unstable}$^*$&  {Stable}$^*$&1\\
 & $\frac{\pi}{8},15\frac{\pi}{8}$& {Unstable}&  {Stable}$^*$&$16$\\
 & $\frac{\pi}{4},7\frac{\pi}{4}$& {Unstable}&  Unstable&$8$\\
 & $3\frac{\pi}{8},13\frac{\pi}{8}$& {Stable}&  {Unstable}&$16$\\
 & $\frac{\pi}{2},3\frac{\pi}{2}$& {Stable}$^*$&  {Unstable}&$4$ \\
 & $5\frac{\pi}{8},11\frac{\pi}{8}$& {Unstable}&  {Unstable}&$16$\\
 & $3\frac{\pi}{4},5\frac{\pi}{4}$& {Stable}$^*$&  {Unstable}&$8$\\
 & $7\frac{\pi}{8},9\frac{\pi}{8}$& {Stable}&  {Unstable}&$16$\\
 & $\pi$& {Stable}$^*$&  {Unstable}&$2$\\\hline
 % \end{longtable}
 % \end{tabular}
% \end{table}

%   \begin{table}
%        \centering
%   \caption{Phase model predictions across different values of $g_M=\{0,2,5\}$ in RTM networks equipped with {\textit{distance-dependent coupling with $p = 0.5$.}}. \textit{Continuation from Table \ref{tab:AA DD 0.5 psi stable}} } 
% \renewcommand{\arraystretch}{1}
     % \begin{tabular}{cccccc}\hline 
     % Network Size&Phase Difference&\multicolumn{3}{c}{${g_M}$} & Cluster Size\\
      
     % \cline{3-5} $N$ &$\mathbf{\psi}$& ${g_M = 0 }$& ${g_M = 2 }$& ${g_M = 5 }$& ${n}$\\\hline
 \multirow{12}*{24}& $0$& {Unstable}$^*$&  {Stable}$^*$&1\\
 & $\frac{\pi}{12},23\frac{\pi}{12}$& {Unstable}&  {Stable}$^*$&$24$\\
      &$\frac{\pi}{6},11\frac{\pi}{6}$&   {Unstable}& {{Unstable}}& $12$\\
 & $\frac{\pi}{4},7\frac{\pi}{4}$& {Unstable}&  {Unstable}&$8$\\
          &$\frac{\pi}{3},5\frac{\pi}{3}$& {Stable}&{Unstable}& $6$ \\
 & $5\frac{\pi}{12},19\frac{\pi}{12}$& {Stable}&  {Unstable}&$24$\\
          &$\frac{\pi}{2},3\frac{\pi}{2}$& {Stable}$^*$&{Unstable}& $4$ \\
 & $7\frac{\pi}{12},11\frac{\pi}{12}$& {Stable}&  {Unstable}&$24$\\
          &$5\frac{\pi}{6},7\frac{\pi}{6}$& {Stable}& {Unstable}& $12$ \\
          &$2\frac{\pi}{3},4\frac{\pi}{3}$& {Stable}$^*$& {Unstable}& $3$ \\
 & $3\frac{\pi}{4},5\frac{\pi}{4}$& {Stable}&  {Unstable}&$8$\\
          &$\pi$&  {Stable}$^*$&  {Unstable}&$2$\\\hline
    \end{tabular}
    
    \caption{Phase model predictions $g_M =0$ (high ACh) and $g_M=5$ (low ACh) in RTM networks equipped with \textit{distance-dependent} coupling with $p = 0.5$. The solutions that were verified through numerical simulation of the full network model are marked with a $^*$.}\label{tab:AA DD 0.5 psi stable}\label{tab:AA DD 24 0.5 psi stable}
    \end{table}
\subsubsection{Networks with \textit{nearest neighbour} coupling}    
As our final network example, take the case where each neuron is only connected to its nearest neighbours. Note that this is equivalent to taking $p=0$ in the distance dependent case. As for distance dependent coupling, for a network of size $N$ the admissible symmetric cluster solutions are ones where adjacent cells are spiking with a phase difference $\psi=\frac{2\pi k}{N}$ for some $k \in \{1,2,..,N\}$. Recall from sections~\ref{sec:NNstab} that the stability of these solutions is determined by the sign of $H_{odd}'(\psi)$. Thus stability does not depend {\em explicitly} on the size of the network. $N$ only determines which phase differences correspond to admissible solutions.
The phase model predictions are summarized in Table \ref{tab:psi stable} for cluster sizes $n=1,2,..,10,12,15$.  Each row in Table~\ref{tab:psi stable} shows the  stability for $\psi$ corresponding to a particular $n-$cluster solution, for the two choices of M-current conductance strength, $g_M$. The synchronized ($1-$cluster) solution, as expected, follows the same stability trend as the prior cases, with the synchronized solution being stable for $g_M=5$. Unlike the previous cases, when $g_M$ is set to $5$, the phase model predicts that there are multiple stable $n-$cluster solutions beyond the splay state and the synchronized state. When $g_M$ is set to $0$, the phase model predictions are similar to the previous cases, with multiple stable $n-$cluster solutions. Of particular note, the $4-$, $5-$ and $9-$ cluster solutions are predicted to be stable for both values of $g_M$ with nearest neighbour coupling, while with the other coupling schemes they were predicted to be stable only at $g_M=0$, if at all.
Note that the predictions made in Table~\ref{tab:psi stable} are independent of $N$. So the table predicts the stability of all symmetric cluster solutions for networks of all sizes up to $N=15$. For a larger network, it predicts the stability of a subset of the symmetric cluster solutions. For example with $N=24$, the table predicts the stability of  all possible symmetric cluster solutions except the $24-$cluster (splay) solutions. 
\begin{table}[h!]
    \centering
     \renewcommand{\arraystretch}{1}
     \begin{tabular}{cccc}\hline
     Phase Difference &\multicolumn{2}{c}{Value of $g_M$}& Cluster Size\\
   ${\psi}$& ${0 }$ & ${ 5 }$&  ${n}$\\ \hline \hline
         $0$&  Unstable$^*$&{{Stable}$^*$}&1\\
 $2\frac{\pi}{15},28\frac{\pi}{15}$& Unstable&  {{Stable}$^*$}&$15$\\
     $\frac{\pi}{6},11\frac{\pi}{6}$&   Unstable& {{Stable}$^*$}& $12$\\
         $\frac{\pi}{5},9\frac{\pi}{5}$& Unstable&{{Stable}$^*$}& $10$\\
 $2\frac{\pi}{9},16\frac{\pi}{9}$& Unstable&  {Stable}&$9$ \\
         $\frac{\pi}{4},7\frac{\pi}{4}$&  Unstable& {{Stable}$^*$}& $8$ \\
 $2\frac{\pi}{7},12\frac{\pi}{7}$& Unstable&  {{Stable}$^*$}&$7$\\
 $4\frac{\pi}{15},26\frac{\pi}{15}$& Unstable&  {{Stable}$^*$}&$15$\\
         $\frac{\pi}{3},5\frac{\pi}{3}$& Unstable&{{Stable}$^*$}& $6$ \\
 $2\frac{\pi}{5},8\frac{\pi}{5}$& {{Stable}$^*$}&  {{Stable}$^*$}&$5$ \\
 $4\frac{\pi}{9},14\frac{\pi}{9}$& {Stable}&  {Stable}&$9$ \\
 $\frac{\pi}{2},3\frac{\pi}{2}$& {{Stable}$^*$}&  {Stable}&$4$ \\
 $4\frac{\pi}{7},10\frac{\pi}{7}$& {{Stable}}&  Unstable&$7$\\
 $8\frac{\pi}{15},22\frac{\pi}{15}$& {Stable}&  Unstable&$15$\\
 $3\frac{\pi}{5},7\frac{\pi}{5}$& {Stable}&  Unstable&$10$ \\
          $4\frac{\pi}{5},6\frac{\pi}{5}$& {{Stable}$^*$}& Unstable& $5$ \\
 $6\frac{\pi}{7},8\frac{\pi}{7}$& {{Stable}$^*$}&  Unstable&$7$\\
         $5\frac{\pi}{6},7\frac{\pi}{6}$& {Stable}& Unstable& $12$ \\
         $3\frac{\pi}{4},5\frac{\pi}{4}$& {Stable}& Unstable& $8$ \\
 $2\frac{\pi}{3},4\frac{\pi}{3}$& {{Stable}$^*$}&  Unstable&$3$ \\
 $8\frac{\pi}{9},10\frac{\pi}{9}$& {{Stable}$^*$}&  Unstable&$9$\\
 $14\frac{\pi}{15},16\frac{\pi}{15}$& {{Stable}$^*$}&  Unstable&$15$\\
         $\pi$&  {{Stable}$^*$}&   Unstable$^*$&$2$\\
          \hline
    \end{tabular}
    \caption{Phase model predictions for $g_M =0$ (high ACh) and $g_M=5$ (low ACh) in RTM networks with \textit{nearest neighbour} coupling. The solutions that were verified through numerical simulation of the full network model are marked with a $^*$. }
    \label{tab:psi stable}
\end{table}
\subsection{Numerical Simulations}\label{initial cond}
To test the phase model predictions, we carried out numerical simulations of the full model equations Eqs.  \eqref{K&O}--\eqref{w_eq}. Predictions that were verified by numerical simulations are marked with an * in Tables \ref{tab:psi stable Glob Hom}--\ref{tab:psi stable}.  All model simulations were carried out using a 2nd-order Runge Kutta algorithm (ODE23) in MATLAB using default numerical parameters \parencite{doi:10.1137/S1064827594276424}. The model parameter values are as described in Section \ref{MOdel}. The number of time steps for all simulations was set to $1.5\times 10^4$, with two exceptions. To satisfy the weak coupling requirement, we set $g_{syn}$ in Eqs.  \eqref{K&O}--\eqref{w_eq} to $0.2$.
% First, in the case of a pair of weakly coupled RTM model neurons, the time steps were increased to $3 \times 10^4$. 
Below, we present the results for two network configurations differing in their coupling architecture, and compare the stability at two levels of the M-current.

\subsubsection{Networks of with all-to-all \textit{globally homogeneous} coupling}
\begin{figure}
    \centering
    \includegraphics[width=0.99\linewidth]{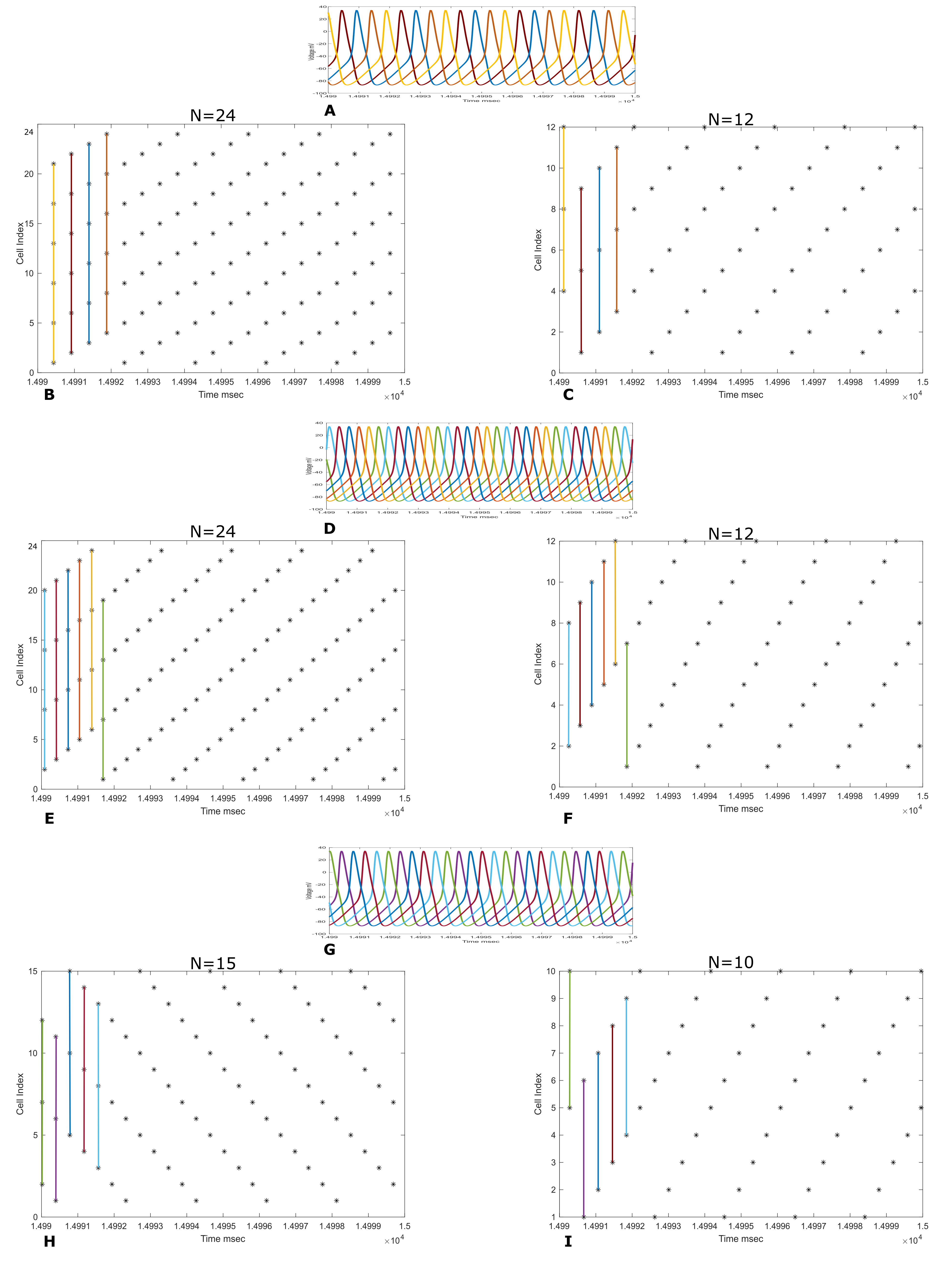}
    \caption{\footnotesize  Multiple stable cluster solutions obtained for a network of reduced Traub-Miles model neurons of sizes $N=10,12,15$ and $24$, with global homogeneous coupling and $g_M=0$ (high ACh).
    %In the network, cells are coupled to every other cell with equal synaptic weights ($k=\lfloor \frac{N}{2}\rfloor$,$p=1$) with $g_M=0$. 
    The voltage traces are presented for a \textbf{(A)} $4-$cluster, \textbf{(D)} $6-$cluster and \textbf{(G)} $5-$cluster solutions. A stable $4-$cluster solution is illustrated for a network of \textbf{(B)}  $24$ cells and   \textbf{(C)} $12$ cells, both associated with $\psi=\frac{\pi}{2}$.  A stable $6-$cluster is illustrated in a network of \textbf{(E)} $24$ cells and \textbf{(F)} $12$ cells, both associated with $\psi=\frac{\pi}{3}$. \textbf{(H)}  A stable $5-$cluster solution is illustrated for a network of  $15$ cells associated with $\psi=\frac{2\pi}{5}$. \textbf{(I)}A different stable $5-$cluster is illustrated for a network of $10$ cells with $\psi=\frac{8\pi}{5}$. The order of cluster firing is discussed in the main text.}
    \label{fig:SimAllGhGm0}
\end{figure}

When $g_M$ is set to $0 $,  multiple $n-$cluster solutions were predicted by the phase model to be stable for a network of any size with globally homogeneous all-to-all coupling.  In Figure  \ref{fig:SimAllGhGm0}, panels (A), (D) and (G) display the voltage traces corresponding to the $4-,6-$ and $5-$ cluster solutions that were obtained in the numerical simulations for a network of 10, 12, 15 and 24 RTM model cells. The remaining panels, (B),(C),(E),(F),(H) \& (I) below each voltage trace illustrate how a network organizes itself into the $n$ clusters. For example, consider panel (B), which shows a network of $24$ RTM model neurons equipped with all-to-all connections exhibiting a stable $4-$cluster solution. The raster plots portray the evolution of the clusters as a time series.  We can determine the sequence in which each of the $4$ clusters spike in one cycle.  In this case, the $4$ clusters for the network of $24$ neurons are given by:
\begin{align*}
    & C_1 = \{1,5,9,..,21\}
    &\hspace{-.2cm} C_2 = \{2,6,10,..,22\} \hspace{0.2cm}
    & C_3 = \{3,7,11,..,23\}
    &\hspace{-.2cm} C_4 = \{4,8,12,..,24\}
\end{align*}
In panel (B), the raster plot indicates that the firing order is $C_1-C_2-C_3-C_4$ with $\psi =\frac{\pi}{2}$. The same type of solution $\left(\psi=\frac{\pi}{2}\right)$ is also observed in a network of size $N=12$, shown in panel (C). For $\psi=\frac{3\pi}{2}$, the clusters would be the same but the firing order would be reversed $C_4-C_3-C_2-C_1$.  Now consider the $6-$cluster solutions obtained for $N=24$ illustrated in panel (E). The clusters are given by:
\begin{align*}
    & C_i = \{i,i+6,i+2\cdot 6,i+3\cdot 6\}.
\end{align*}
Similarly the $6-$ cluster solutions obtained for $N=24$ illustrated in panel (E) are ordered  $C_1-C_2-C_3-C_4-C_5-C_6$ corresponding to $\psi=\frac{\pi}{3}$. The same type of solution $\left(\psi=\frac{\pi}{3}\right)$ is also observed for a $12$ neuron network illustrated in panel (F). We were unable to find the other $4-$ and $6-$ cluster solutions predicted in Table~\ref{tab:psi stable Glob Hom} through numerical simulation.

An example of a $5-$cluster with different ordering is illustrated in panels (H) and (I) for a network of $15$ and $10$ RTM model neurons, respectively. Following the same notation as above, the clusters are given by: $C_i = \{i,i+5,..,N-n-i\}$ for $i=\{1,2,3,4,5\}$. In panel (H), the clusters fire in the order of $C_5-C_4-C_3-C_2-C_1$ corresponding to a phase difference of $\psi=\frac{2\pi}{5}$. The ordering is reversed in the case in (I), with an ordering of $C_1-C_2-C_3-C_4-C_5$ corresponding to $\psi=\frac{8\pi}{5}$. The colours are assigned in order to match the order of the voltage traces presented in panel (G). Note that the $5-$cluster solution can also take on other orderings as well, such as those associated with $\psi=\frac{4\pi}{5}$ ($C_1-C_4-C_2-C_4-C_3$) and $\psi=\frac{6\pi}{5}$ ($C_1-C_3-C_5-C_2-C_4$) which we did not find through numerical simulation.

\begin{figure}[h!]
    \centering
    \includegraphics[width=\linewidth]{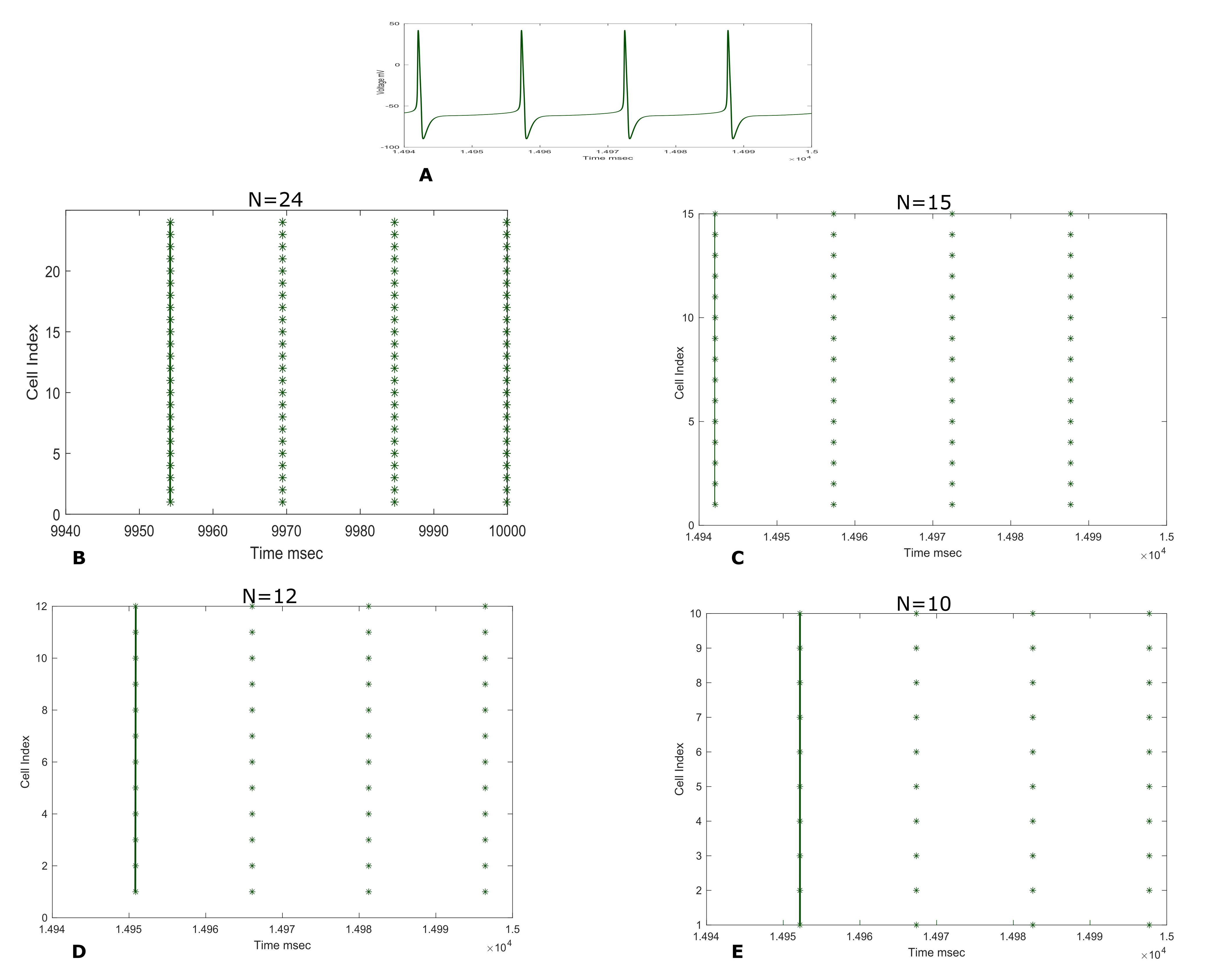}
    \caption{Synchronization of a network of RTM model neurons with global homogeneous coupling, when $g_M=5$ (low ACh). \textbf{(A)} The voltage traces associated for the synchronized ($1-$cluster) solution. The synchronized solution is illustrated for a network of \textbf{(B)} $24$ cells, \textbf{(C)} $15$ cells, \textbf{(D)} $12$ cells and \textbf{(E)} $10$ cells through raster plots.}
    \label{fig:SimAllGhGm5}
\end{figure}

Increasing the $g_M$ to $5$, the system with global homogeneous coupling strongly tends to the synchronized solution, for all network sizes we explored. Other states were verified to be unstable by initializing the system close to the cluster solution.
% \footnote{In this case we confirmed that an $n-$cluster corresponding to one $\psi$ is unstable, we did not check it for all values of $\psi$} . 
The synchronization of the network is demonstrated in Figure \ref{fig:SimAllGhGm5}.


\subsubsection{Networks with all-to-all \textit{distance-dependent} coupling}
We studied networks with $N=$ $4,5,\ldots,16$ and $24$  equipped with distance dependent coupling where $p=0.5$ in Eq.~\eqref{mujDDeve} or \eqref{mujDDodd}. When $g_M=0$, we obtained multiple stable $n-$cluster solutions for each $N$, in agreement with the phase model predictions. In particular, we obtained the $2-$cluster and $3-$cluster solutions for $N=6, 8, 16$ and $24$. The phase model also predicts that an $8-$cluster solution corresponding to $\psi=3\frac{\pi}{4}$ (or $5\frac{\pi}{4}$) is stable when $g_M=0$ for $N=8$  and $16$. We obtained these solutions for both networks. Simulations showing the $2-$cluster, $3-$cluster and $8-$cluster solutions obtained for $g_M=0$ are presented and discussed in \parencite{manoj_phase_2025}. 

\begin{figure}[h!]
    \centering
    \includegraphics[width=\linewidth]{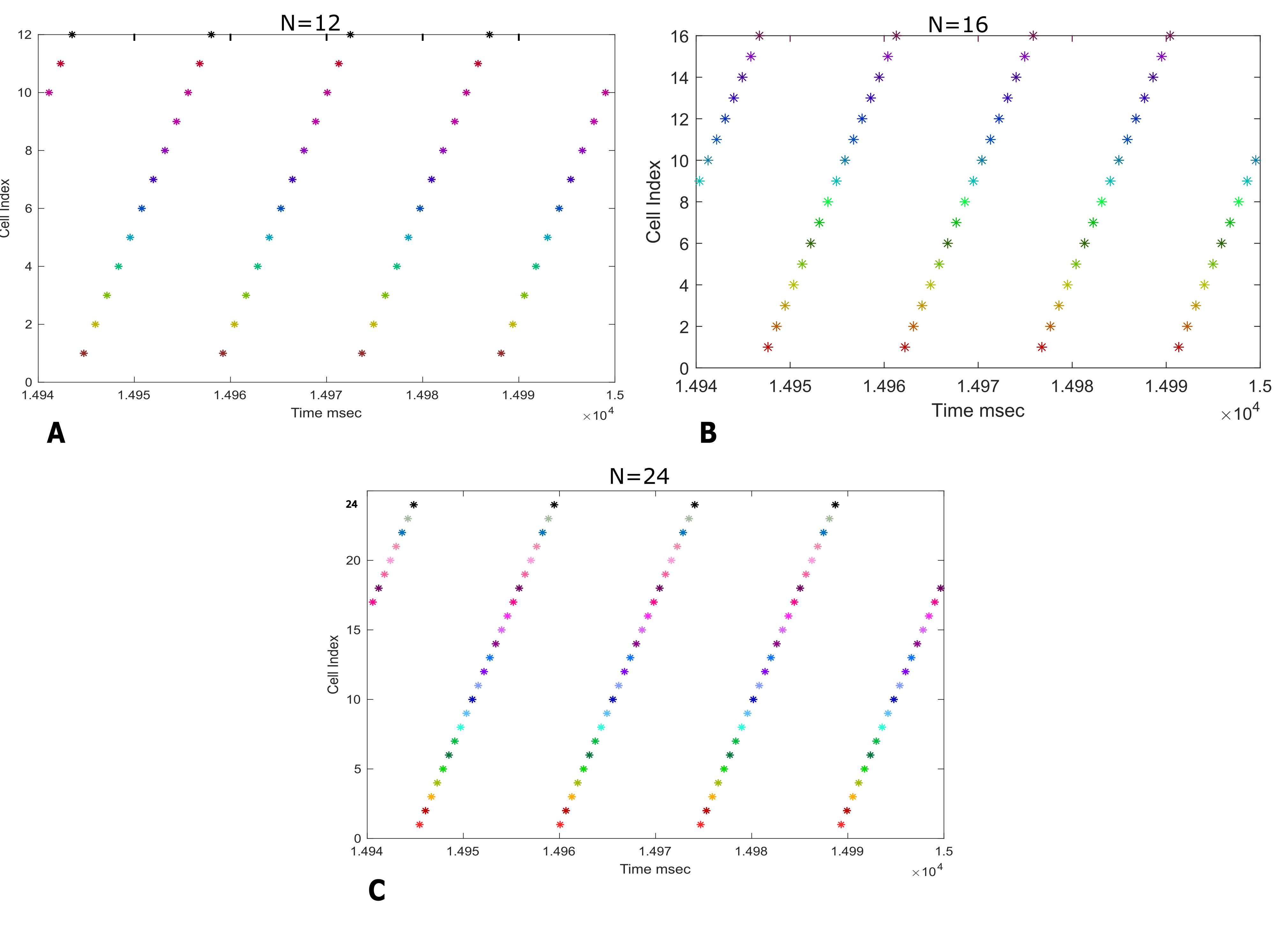}
    \caption{\footnotesize Demonstration of splay state solutions for RTM networks equipped with distance-dependent coupling ($p=0.5,k=\lfloor N/2\rfloor$) with $g_M=5$ (low ACh).  \textbf{(A)} A splay solution for a network of $12$ cells, spiking with a phase difference $\psi=\frac{\pi}{6}$.  \textbf{(B)} A splay solution for $16$ cells with a phase difference $\psi=\frac{\pi}{8}$.  \textbf{(C)} A splay solution obtained for a network of $24$ neurons with $\psi=\frac{\pi}{12}$.  In each case, the cells fire in the order of $1-2-3-\dots-N$ as illustrated through the raster plots. }
    \label{fig:AADDgm5}
\end{figure}

 When $g_M$ is increased to $5$, networks with distance-dependent coupling fully synchronize, as in the case with globally homogeneous coupling. For $N>8$, the phase model also predicts an additional stable splay solution for the network. In particular, it predicts a splay state in which adjacent cells  spike with a phase difference of  $\psi = 2\frac{\pi}{N}$ or $(2N-2)\frac{\pi}{N}$.  These type of solutions were obtained for $N=12,16$ and $24$ and are presented in Figure \ref{fig:AADDgm5}. In each case illustrated in panels (A)-(C), cells within the network, spike in the order of $C_1-C_2-\dots-C_{N-1}-C_N$ corresponding to a phase difference of $\psi=2\frac{\pi}{N}$ for $N=12,16$ and $24$ respectively.

\subsubsection{Networks with \textit{nearest neighbours} coupling.}
We studied network with nearest neighbours coupling,  $N=5,6,\ldots,15$. 
% and $g_{syn}=0.2$. 
In general, the symmetric cluster solutions that emerged in the simulations agreed with the phase model predictions.  When $g_M=0$, we found multiple $n-$cluster solutions consistent with those predicted to be stable by the phase model. In particular, we obtained the $2-$cluster, $3-$cluster and $5-$cluster solutions for all network sizes that we explored. Although the phase model predicts stable $4-$cluster, $6-$cluster and $8-$cluster solutions, we were unable to  obtain these solutions for an network with $N>4$ even. We did, however, obtain the  $4-$cluster for a network of $4$ cells. In addition to this, we obtained splay states for $N=7$ and $9$. These solutions are presented and discussed in \parencite{manoj_phase_2025}. 

\begin{figure}
    \centering
    \includegraphics[width=\linewidth]{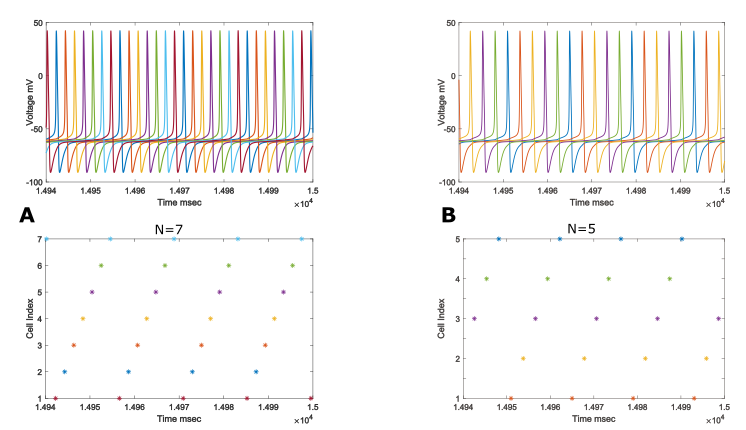}
    \caption{Splay state solutions for a network of RTM model neurons, equipped with nearest neighbour coupling, with $g_M=5$ (low ACh).    \textbf{(A)} A network of $7$ cells exhibiting a $7-$cluster solution with $\psi=\frac{12\pi}{7}$. \textbf{(B)} A network of $5$ cells exhibiting a $5-$cluster solution with $\psi=\frac{2\pi}{5}$.  }
    %\caption{Splay state solutions for a network of RTM model neurons, equipped with nearest neighbour coupling, with $g_M=5$.    \textbf{(A)} A network of $7$ cells exhibiting a $7-$cluster solution with $\psi=\frac{12\pi}{7}$. \textbf{(B)} A network of $5$ cells spike individually corresponding to a $5-$cluster solution with $\psi=\frac{2\pi}{5}$.  }
    \label{fig:splaygm5}
\end{figure}

When $g_M=5$, in addition to the fully synchronized solution ($\psi=0$, not illustrated), we also found several non-trivial $n-$cluster solutions\footnote{By non-trivial, we mean $n-$cluster solutions with $n>1$.}. Unlike the distance dependent case, we could find splay states for networks smaller than $10$ cells. As an example, consider Figure \ref{fig:splaygm5}, where panels (A) and (B) show the splay states that were obtained for a network of $7$ and $5$ cells, associated with $\psi=\frac{12\pi}{7} (C_1-C_7-C_6-C_5-C_4-C_3-C_2)$ and $\psi=\frac{2\pi}{5} (C_1-C_2-C_3-C_4-C_5)$ respectively. 

\begin{figure}
    \centering
    \includegraphics[width=1\linewidth]{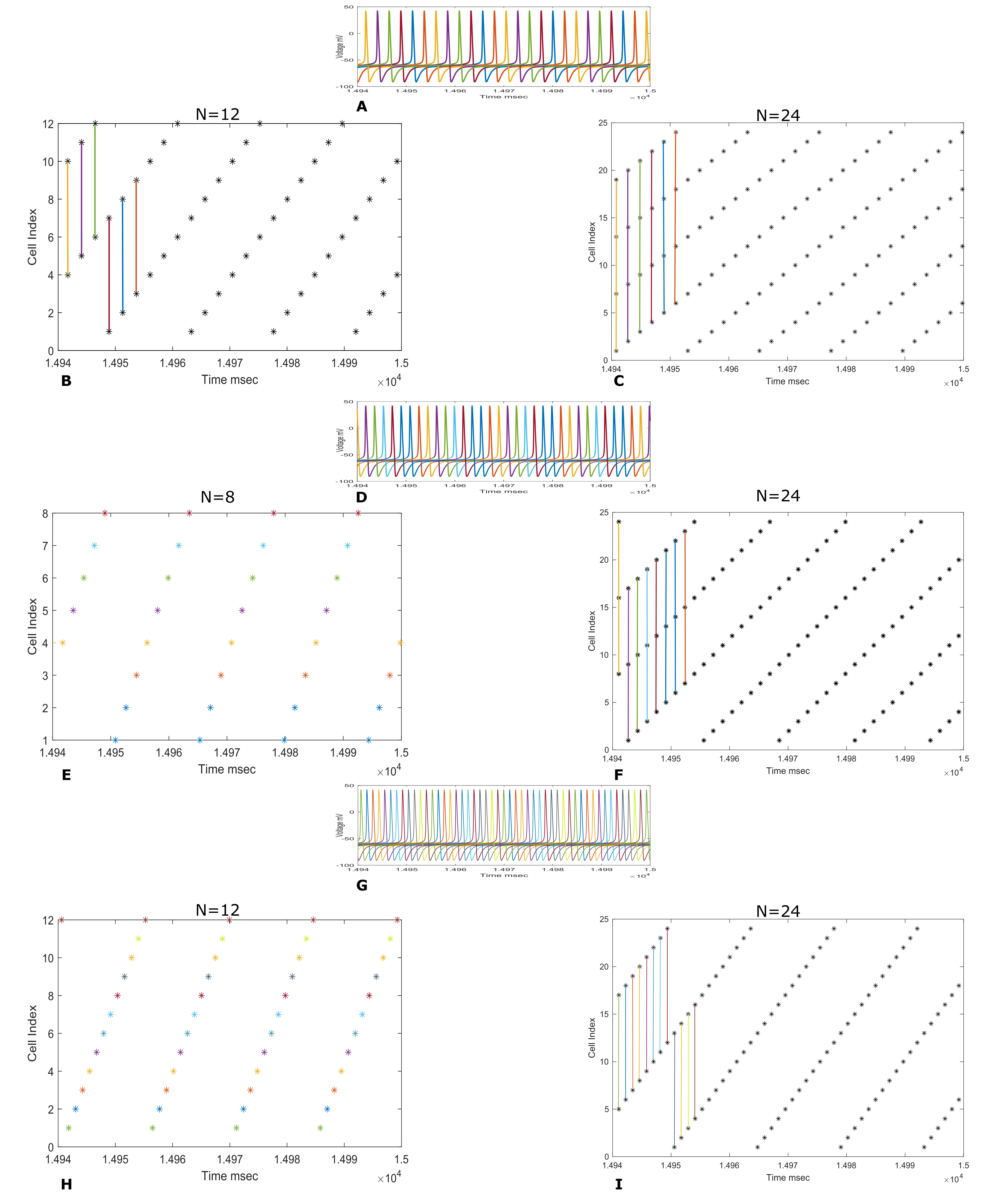}
    \caption{Multiple stable $n-$cluster solutions are obtained for a network RTM model neurons with nearest neighbour coupling and $g_M=5$ (low ACh). \textbf{(A),(D)\& (G)} display the representative voltage traces for each solution. Panels \textbf{(B)} and \textbf{(C)} illustrate the $6-$cluster solution in a network of $12$ and 24 cells resp, with $\psi=\frac{\pi}{3}$. Panels \textbf{(E)} and \textbf{(F)} demonstrate the $8-$cluster solution in a network of $8$ and $24$ cells resp with $\psi=\frac{\pi}{4}$. Panels\textbf{ (H)} and \textbf{(I)} illustrate the $12-$cluster in a network of $12$ and $24$ cells resp, with $\psi=\frac
    {\pi}{6}$}
    \label{fig:nclusgm5}
\end{figure}

Further splay solutions for networks with $N=8$ and $12$ cells are shown in Figure \ref{fig:nclusgm5}. These correspond to $\psi=\frac{2\pi}{N}$ and the firing order of cells is $C_1-C_2-\dots-C_N$.  Figure \ref{fig:nclusgm5} also shows non-trivial cluster solutions that were not splay states that we obtained. Panels (G),(H) and (I) show $6-$cluster, $8-$cluster and $12-$cluster solutions in a network of $24$ cells. Panel (D) shows a $6-$cluster solution in a network of $12$ cells. In all cases, the cells fire in the same order, i.e., $\{C_1-C_2-\dots C_N\}$ corresponding to $\psi=\frac{2\pi}{N}$ with $N=12 $ or $ 24$ accordingly.
 
\subsubsection{Persistence of solutions under transient perturbations to the applied current}
According to the phase model analysis, RTM networks equipped with a fixed synaptic architecture can exhibit multiple stable cluster solutions for some fixed level of the M-current. This suggests potential pathways or transitions between different stable cluster solutions within the same system. One may obtain these transitions by introducing transient perturbations to a system that has already reached a stable cluster. In this paper, we use transient perturbations in the form of an increased external current pulse applied during a short time interval, $[8000,8050]$ ms. The corresponding system that has been initialized in a predicted stable solution, can respond in one of two ways to this perturbation: the solution persists or will transition into another distinct stable solution.

First, we considered a network with globally homogeneous coupling and no M-current ($g_M=0$), initialized in a stable cluster solution, and introduced transient perturbations to subset of the cells. The perturbations could switch the network between different realizations of the same $n-$clusters. This could either be a change to the firing order, or a change in the cells that belong to the clusters. We present a few of these transitions for the networks  $N=12,15$ and $24$. 

\begin{figure}[h!]
    \centering
    \includegraphics[width=\linewidth]{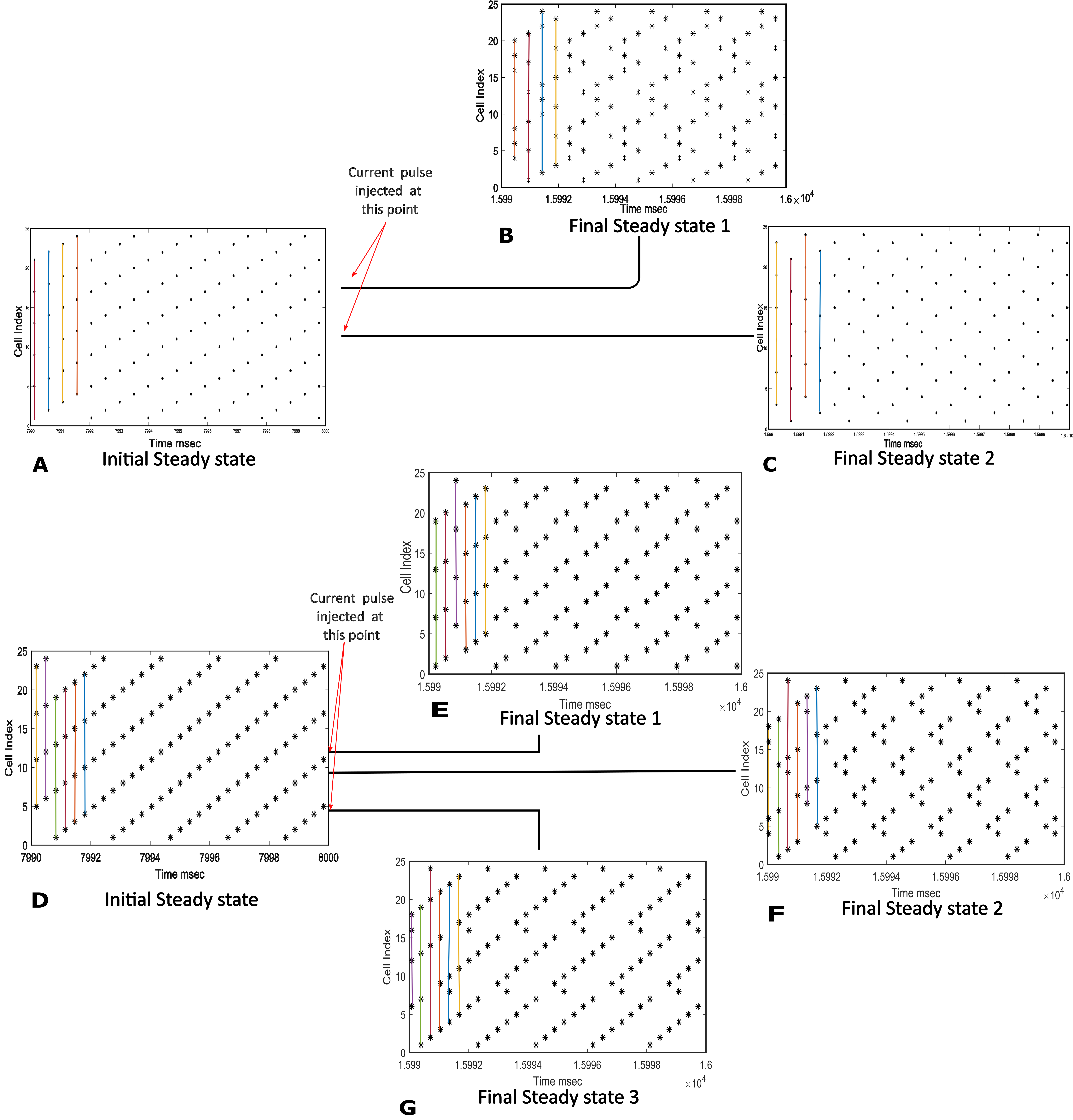}
    \caption{\footnotesize Multistability in a network of $24$ RTM model neurons with globally homogeneous coupling and $g_M=0$ (high ACh).  \textbf{(A)} The network is initialized in a stable $4-$cluster with a firing order: $C_1-C_2-C_3-C_4$, where $C_i=\{i,i+4,\dots,19+i\}$. \textbf{(B)} A transient perturbation, in the form a current pulse applied to a subset of cells, can switch the network to a stable $4-$cluster solution where the cells belonging to $C_2$ and $C_4$ change to $\{2,10,12,14,22,24\}$ and $\{4,6,8,16,18,20\}$, respectively. \textbf{(C)}  Applying the pulse to a different subset of the cells alters the firing order to $C_1-C_4-C_2-C_3$. \textbf{(D)} The network is initialized in a stable $6-$cluster with a firing order: $C_1-C_2-C_3-C_4-C_5-C_6$, where $C_i=\{i,i+6,\dots,17+i\}$. \textbf{(E)} A transient perturbation, in the form a current pulse applied to subset of the cells, alters the firing order to  $C_1-C_2-C_6-C_3-C_4-C_5$.  \textbf{(F)} The perturbation can also switch the network to a stable $6-$cluster solution where the cells belonging to cluster $C_6$, $C_2$ and $C_4$ change to $\{8,10,20,22\}$,$\{2,12,14,24\}$ and $\{4,6,16,18\}$ and the firing order changes to $C_1-C_2-C_3-C_6-C_5-C_4$.    \textbf{(G)}  Applying a perturbation to a different subset of cells can also switch the network to a stable $6-$cluster solution where the cells belonging to cluster $C_6$, $C_2$ and $C_4$ change to $\{6,11,16,18\}$,$\{2,13,20,24\}$ and $\{4,8,10,22\}$.    }
    \label{fig:AAGHtrans1524}
\end{figure}

In Figure \ref{fig:AAGHtrans1524}, panel (A) shows a network of $24$ RTM model neurons, initialized in a stable $4-$cluster associated with $\psi=\frac{\pi}{2}$ and a firing order given by $C_1-C_2-C_3-C_4$.  Applying a transient pulse to a subset of the cells can switch the network to a new stable $4-$cluster solution, either with a different ordering given by: $C_1-C_4-C_2-C_3$, as shown in panel (C), or a different arrangement of cluster cells as shown in panel (C). In panel (A), the solution is initialized with clusters, $C_2$ and $C_4$, given by: $\{2,6,10,14,18,22 \}$ and $\{4,8,12,16,20,24\}$ respectively.  In panel (B), clusters $C_2$ and $C_4$ recruit different cells in the new stable solution given by: $\{2,10,12,14,22,24 \}$ and $\{4,6,8,16,18,20\}$ respectively.

Now consider the same network initialized in a stable $6-$cluster solution with firing order given by: $C_1-C_2-C_3-C_4-C_5-C_6$ which is displayed in panel (D) of Figure \ref{fig:AAGHtrans1524}. A transient perturbation, applied to a subset of the cells,  can switch the initial state to a new stable cluster solution in three ways as demonstrated through panels (E)-(G). Panel (E) is a stable $6-$cluster solution with a different firing order given by: $C_1-C_2-C_6-C_3-C_4-C_5$. Panel (G) shows a stable $6-$cluster solution where the clusters formed in the initial state (panel (D)) change. In panel (G), the cluster $C_6$ changes from $\{6,12,18,24\}$ to $\{6,12,16,17\}$, the cluster $C_2$ changes from $\{2,8,14,20\}$ to $\{2,13,20,24\}$, and the cluster $C_4$ changes from $\{4,10,16,22\}$ to $\{4,8,10,22\}$.  Panel (F) shows a stable $6-$cluster solution where there is a change to both the firing order and the arrangement of cells in the clusters.  In panel (F), the clusters $C_4,C_2$ and $C_6$ change to $\{4,6,16,18\}$, $\{2,12,14,24\}$ and $\{8,10,20,22\}$ respectively. The firing order is given by: $C_1-C_2-C_3-C_6-C_5-C_4$. A description of the results associated with $N=12$ and $15$ are discussed in \parencite{manoj_phase_2025}.

The simulations were repeated for the same network with $g_M=5$, where we observed that the synchronized solution persisted under the same transient current pulses used in the $g_M=0$ case.

\begin{figure}[h!]
    \centering
    \includegraphics[width=0.95\linewidth]{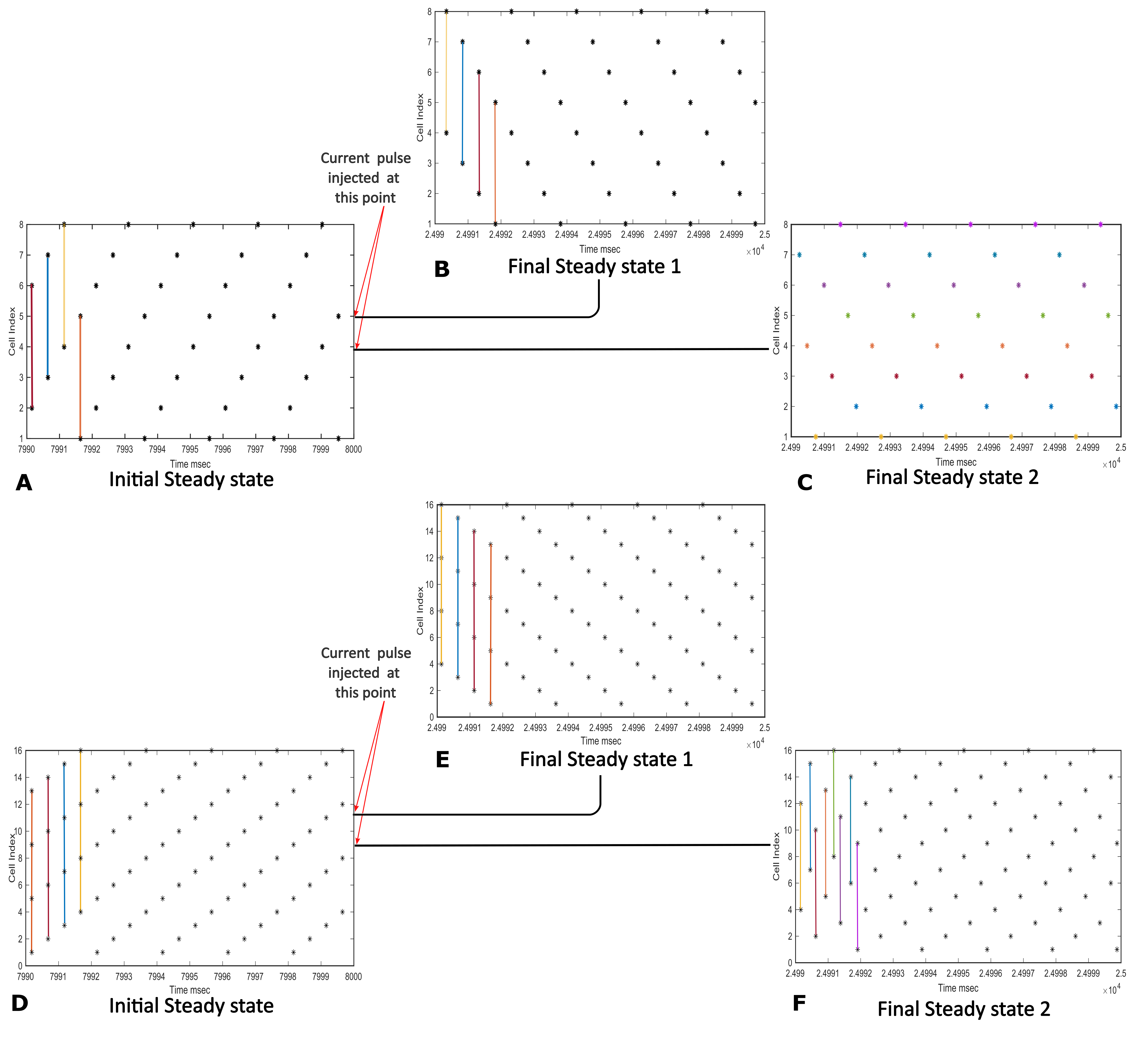}
    \caption{\footnotesize Multistability in a network of RTM model neurons, equipped with global distance dependent coupling ($p=0.5$), when $g_M=0$ (high ACh). In each case, the network is initialized in a stable cluster solution and a transient perturbation is applied to a subset of the cells during time $t\in[8000,8050]$. \textbf{(A)} Network of $8$ cells is initialized in a stable $4-$cluster solution $\left(\psi=\frac{\pi}{2}\right)$. \textbf{(B)} The perturbation reverses the firing order of the spiking cells from (A), switching the network to a different stable $4-$cluster solution $\left(\psi=3\frac{\pi}{2}\right)$. \textbf{(C)} A  perturbation applied to a different subset of cells, can switch the initial state in (A) to a stable $8-$cluster  $\left(\psi=7\frac{\pi}{4}\right)$. \textbf{(D)} Network of $24$ cells is initialized in a stable $4-$cluster solution $\left(\psi=\frac{\pi}{2}\right)$. \textbf{(E)} The perturbation switches the network in (D)  to a different stable $4-$cluster solution ($\psi=3\frac{\pi}{2}$) \textbf{(F)} The network in (D) can also switch to a stable $8-$cluster $\left(\psi=7\frac{\pi}{4}\right)$.  } 
    \label{fig:AADDTrans1}
\end{figure}

Next, we initialized a network with distance dependent coupling and no $M-$current ($g_M=0$) in one of the predicted stable $n-$cluster solutions and applied transient perturbations to a subset of the cells. The perturbations can switch the network into a different stable $n-$cluster or a different realization of the initial one. In Figure \ref{fig:AADDTrans1}, panels (A) and (D) show the raster plot for a network of $8$ and $16$ cells respectively, each initialized in a stable $4-$ cluster solution corresponding to $\psi=\frac{\pi}{2}$ (with a firing order : $C_1-C_2-C_3-C_4$). Panels (B) and (E) show one case, in which the both networks switch from the initial $4-$cluster solution to a different realization  corresponding to an inter-cluster phase difference of $\psi=3\frac{\pi}{2}$. The final solution is predicted to be a stable solution in Table \ref{tab:AA DD 0.5 psi stable}, and reverses the firing order ($C_1-C_4-C_3-C_2$). Panels (C) and (F) show a second possibility, in which by perturbing a different subset of cells, the network switches to an $8-$cluster solution corresponding to a phase difference $\psi=7\frac{\pi}{4}$ (with a firing order : $C_1- C_6-C_3-C_8-C_5-C_2-C_7-C_4$), which is a reverse of the solution associated with $\psi=3\frac{\pi}{4}$. 

Similarly, for a network of $12$ RTM model neurons with no M-current ($g_M=0$), initialized in a stable $4-$cluster solution with an inter-cluster phase difference of $\psi=\frac{\pi}{2}$,  a transient perturbation can reverse the firing order of the initial state and switch the network to a different stable $4-$cluster solution associated with $\psi=3\frac{\pi}{2}$. The network can also switch into different stable $12-$cluster solutions by applying the perturbation to a different subset of cells, associated with a $\psi$ that is either $5\frac{\pi}{6}$or $7\frac{\pi}{6}$. The solutions for this network is further discussed in \parencite{manoj_phase_2025}. 

\begin{figure}[h!]
    \centering
    \includegraphics[width=1\linewidth]{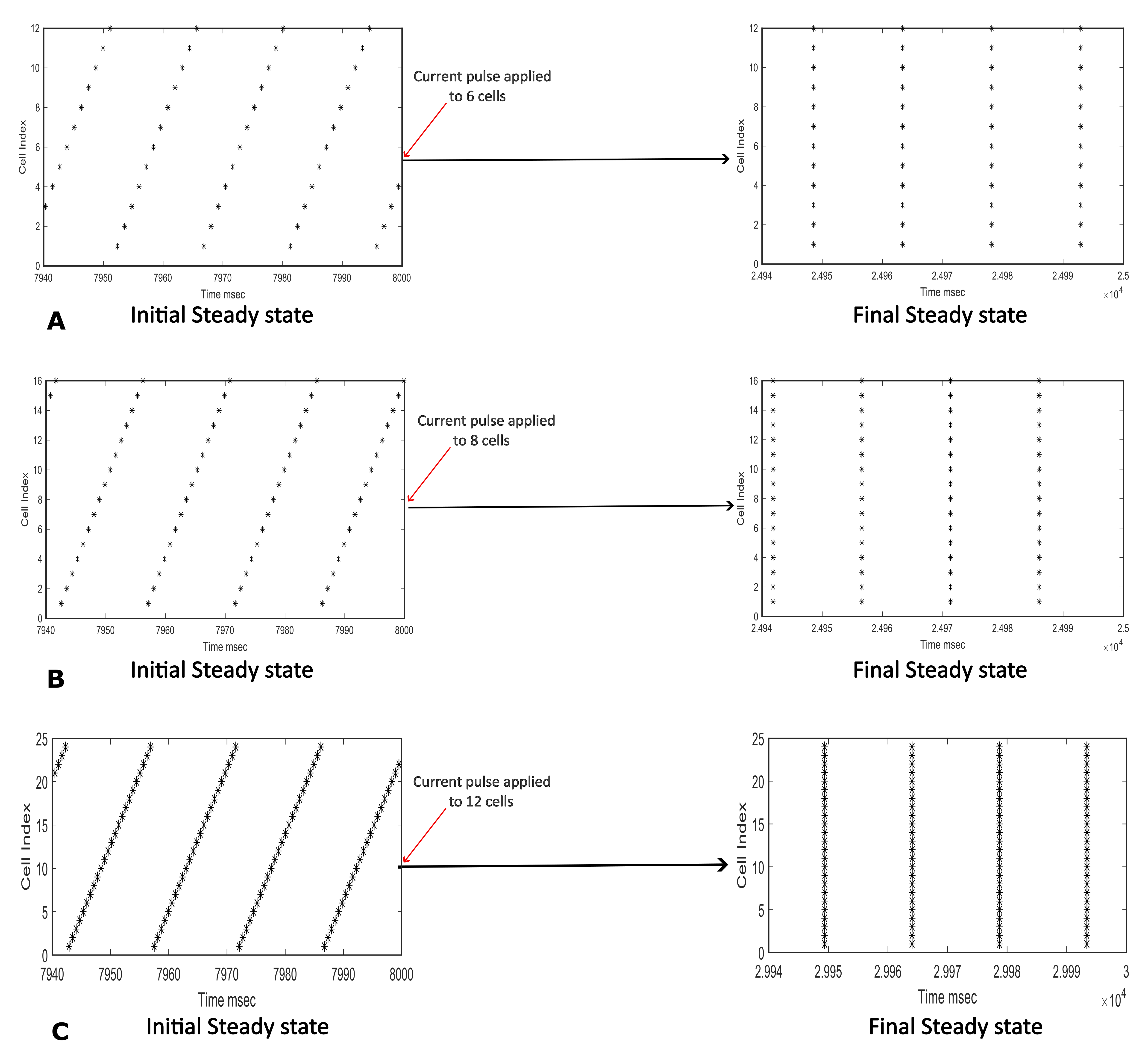}
    \caption{Multistability in a network of RTM model neurons  with distance dependent coupling and $g_M=5$ (low ACh). A transient perturbation is applied during time $t\in[8000,8050]$ to a network initialized in a stable splay solution. \textbf{(A)} Network of $12$ cells, initialized in a stable splay solution $\left(\psi=\frac{\pi}{6}\right)$, synchronizes when $6$ of the cells are perturbed transiently. \textbf{(B)} Network of $16$ cells, initialized in a stable splay solution $\left(\psi=\frac{\pi}{8}\right)$, synchronizes when $8$ of the cells are perturbed transiently. \textbf{(C)} Network of $24$ cells, initialized in a stable splay solution $\left(\psi=\frac{\pi}{12}\right)$, synchronizes when $12$ of the cells are perturbed transiently.}
    \label{fig:splaysynch}
\end{figure}

When $g_M=5$, networks initialized in the synchronized state always returned to this state after the transient perturbations. However, for a network of size $N>8$  (from Table \ref{tab:AA DD 0.5 psi stable}), initialized in a splay solution, applying a perturbation to at least half the cells could synchronize the network.  This is illustrated in panels (A)-(C) in Figure \ref{fig:splaysynch} respectively, for a network of $12$ ,$16$ and $24$ cells , initialized in a splay solution, synchronize after a perturbation is applied to half of the cells. 

Finally, we tested the multistablity of the cluster solutions exhibited in a system with nearest neighbour coupling by applying a transient perturbation, in the form of a current pulse, to a subset of the cells in the network, initialized in one stable cluster solution, during the time interval $[8000,8050]$.  When $g_M=0$, the $2-$cluster solution persists after these current pulses. However, transitions induced by the transient current pulses were observed for networks initialized in a stable $3-$cluster solution. A transient current pulse applied to a subset of the cells could reverse the firing order of the cells forming a stable $3-$cluster solution associated with a different phase difference. Applying the perturbation to a different subset of cells can also switch networks to a stable $12-$cluster solution (if admissible). A third possibility could arise in which a perturbation to a subset of the cells could switch the network to the $2-$cluster solution. These transitions were observed for a network of $12$ and $24$ RTM model neurons, and are presented and discussed in \parencite{manoj_phase_2025}. When $g_M=5$, networks with nearest neighbour coupling initialized in the synchronized state, always returned to this state after the transient perturbations. For networks initialized in a splay state, applying a perturbation to a subset of the cells, could synchronize the solution. This was observed for a network of $8$ and $12$ cells. This pulse-induced synchrony was also observed for a network of $12$ cells that was initialized in a stable $6-$cluster state, associated with either $\psi=\frac{\pi}{6}$ or $\psi=\frac{\pi}{3}$. These results are presented and further discussed in \parencite{manoj_phase_2025}.

\begin{figure}
    \centering
    \includegraphics[width=\linewidth]{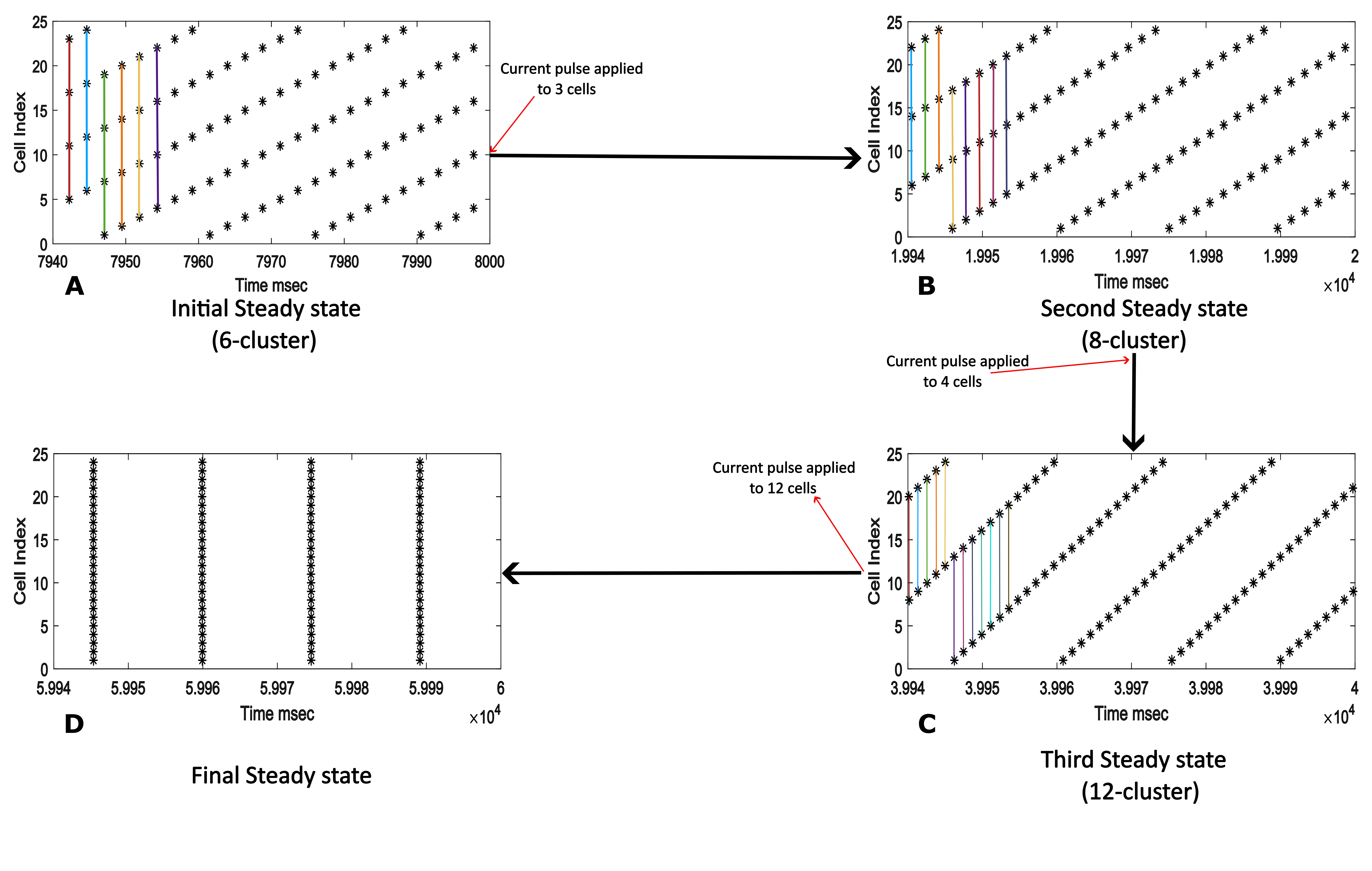}
    \caption{ Demonstration of possible transitions between stable cluster solutions when $g_M=5$ (low ACh), for a network of $24$ RTM model neurons, equipped with nearest neighbour coupling. \textbf{(A)} The network is initialized in a stable $6-$ cluster solution ($\psi=\frac{\pi}{3}$) \textbf{(B)} A transient perturbation applied to 3 cells in the network, can switch the solution in (A) to a stable $8-$cluster solution associated with $\psi=\frac{\pi}{4}$ \textbf{(C)} A transient perturbation applied to 4 cells in the network, can switch the solution in (B) to a stable $12-$cluster solution associated with $\psi=\frac{\pi}{6}$   \textbf{(D)} A transient perturbation applied to 12 cells in the network, can synchronize the network. Transient perturbations applied to  different subset of the cells in the network initialized in (A), (B) or (C), can also synchronize them.}
    \label{fig:NNGM5Trans24}
\end{figure}

The more interesting cases are the possible transitions obtained in a network of $24$ cells, when $g_M=5$, illustrated in Figure \ref{fig:NNGM5Trans24}. Panel (A) depicts the  network initialized in a stable $6-$cluster solution associated with $\psi=\frac{\pi}{3}$. Applying a transient perturbation to 3 cells in the network, can switch the network to a stable $8-$cluster $(\psi=\frac{\pi}{4})$ as depicted in panel (B). And applying a transient current pulse to 4 cells of the network in the second stable (8-cluster) solution can switch it to a stable $12-$cluster solution $(\psi=\frac{\pi}{6})$. Finally, by applying the current pulse to 12 cells of the network in a third stable ($12-$cluster) solution synchronizes the network.  By changing the number of cells perturbed in the first two transitions, we could have induced synchrony either from the initial state (6-cluster) or from the second solution (8-cluster).  By the same logic, we could have also induced a transition directly from the 6-cluster in (A) to the 12-cluster in (C).
\section{Discussion}\label{Discussion}
In this paper, we studied the effect of acetylcholine (ACh) on the synchrony within an excitatory hippocampal network with periodic boundary conditions. We focused on a special type of phase-locked periodic solution of this system, known as a cluster solution. In a cluster solution, all individual cells fire repetitively, and form into subgroups of neurons that fire synchronously, while neurons in different subgroups have fixed, nonzero phase difference.

Acetylcholine is an important neuromodulator that can regulate the dynamics of the hippocampal formation, through which it plays a vital role in cognitive functions, specifically involving learning and memory \parencite{hasselmo_modes_2011,hasselmo_dynamics_1995,hasselmo_laminar_1994}.   To study this effect, we consider a biophysical, conductance-based model of a hippocampal pyramidal neuron, the reduced Traub-Miles (RTM) model \parencite{olufsen_new_2003}. The RTM model includes a voltage dependent non-inactivating slow potassium current 
called  the muscarinic current (M-current), which is known to downregulate in the presence of ACh \parencite{halliwell_m-current_1986, halliwell_voltage-clamp_1982, stiefel_cholinergic_2008}. The corresponding network model satisfied periodic boundary conditions, and had circulant coupling structure. Assuming the coupling between neurons was sufficiently weak, we derived a phase model approximation for this network of RTM model neurons %(with the M-current) 
at two specific strengths of the M-current conductance ($g_M$), linked to low and high levels of ACh, and three different coupling architectures (globally homogeneous, distance dependent, nearest-neighbour). We used the phase model to investigate the influence of ACh on the formation of stable cluster solutions in the network. We specifically focused on symmetric cluster solutions, in which all the clusters contain an identical number of neurons and %neurons organize themselves (temporally) into synchronous clusters, each spiking with fixed inter-cluster phase differences with one another. In this type of solution, 
the phase differences between adjacent neurons are equal 
%and neurons belonging to the same cluster have a phase difference of zero
\footnote{With the exception of all-to-all globally homogeneous coupling, where all $n-$cluster solutions are equivalent, regardless of the phase differences between adjacent neurons.}. Focusing on this type of cluster solution enabled us to give necessary and sufficient conditions for stability of the solutions and hence make meaningful comparisons between different network coupling configurations and levels of ACh.

A network of spiking neurons may %desynchronize, by 
organize into `neural assemblies', which are groups of transiently synchronized neurons. Neural assemblies observed in the hippocampus, are thought to facilitate spatial navigation tasks and the formation of episodic memories \parencite{harris_organization_2003,dragoi_temporal_2006,pastalkova_internally_2008,sakurai_hippocampal_1996}. 
% It is also theorized that, in addition to external inputs, the intrinsic oscillatory dynamics of the hippocampus can also play a role in the formation of neural assemblies \parencite{harris_organization_2003,pastalkova_internally_2008} . 
In our work, the symmetric cluster solutions to biophysical models of neuronal networks represent neural assemblies. A generally accepted consensus, is that high levels of ACh during active exploration and rapid eye movement (REM) sleep, support the encoding of new stimuli, while low levels of ACh during quiet waking and slow wave sleep (SWS), facilitate the consolidation of previously formed memories for future recall \parencite{hasselmo_dynamics_1995,hasselmo_encoding_1996,hasselmo_high_2004, hasselmo_neuromodulation_1999,huang_acetylcholine_2022}. 
We explore this bidirectional role of acetylcholine (ACh),  in neural assembly formation, specifically studying the distinct assemblies that may be supported in our  model network of pyramidal neurons in the hippocampus, under two levels of ACh (high vs. low).

We reviewed phase model results for symmetric cluster solutions for a general network of arbitrary size equipped with nearest-neighbour, %($k=1$) 
 all-to-all globally homogeneous and distance dependent coupling. %($k=\lfloor \frac{N}{2}\rfloor$, $p=1,0.5$). 
%The phase differences % $\psi_i$, between adjacent neurons were identical and denoted by $\psi$.     
We showed that the stability of the synchronized ($1-$cluster) solution was independent of the size of the network %, $N$
and coupling architecture %, $W$, 
and thus, identical to the two neuron case. For symmetric cluster solutions with more than one cluster (nontrivial cluster solutions) the conditions guaranteeing stability depend on the coupling architecture.
%non-trivial solutions ($\psi \neq 0$), the conditions of stability for an $n-$cluster solution, 
In networks with globally homogeneous coupling, the stability depends only on the number of neurons in the cluster. For 
nearest neighbours coupling, the stability depends only on the phase difference between adjacent neurons/clusters. With distance dependent coupling the stability depends on {both} the phase difference and the number of neurons in the network. %\psi$ and $N$.  

% \subsubsection*{\textit{Coupling architecture can modify network susceptibility under low levels of acetylcholine}}
For \textit{low levels of acetylcholine}, %($g_M=5$), 
our results showed a significant difference in the number of stable cluster solutions with the \textit{type of coupling} in %(different $k$ and $p$) that was adopted by 
the network. When  acetylcholine is `low', %($g_M=5$), 
  the only stable solution for networks equipped with all-to-all globally homogeneous coupling, is the synchronized ($1-$cluster) solution.
%arbitrary circulant coupling architectures, could synchronize. As the number of excitatory connections decreases 
%$\left(k:\lfloor \frac{N}{2}\rfloor \rightarrow 1\right)$, 
For nearest-neighbour coupling, the network exhibits multistablity, with several stable symmetric cluster solutions, while still having the ability to synchronize. This suggests that decreasing the number of connections allows for more stable solutions. A similar trend is seen, in which the same number of excitatory connections are maintained (all-to-all)
%$\left(k=\lfloor \frac{N}{2}\rfloor\right)$, 
but the coupling is changed from the globally homogeneous to distance-dependent. Essentially, this means that for a given set of initial membrane voltages, depending on the coupling architecture, the network can either approach a solution with more than one cluster  or it will fully synchronize. This enables the network to respond differentially to a given input, depending on \textit{the number and/or strength of excitatory synaptic connections} under \textit{low levels of ACh}.  % For example, consider a network, under low levels of ACh, equipped with a nearest neighbours or distance dependent coupling, initialized at a non-trivial cluster solution.  For the same set of initial conditions, the corresponding network will synchronize with all-to-all homogeneous excitatory coupling. 
The multistability observed (and numerically verified) in a network  with nearest neighbour or distance-dependent coupling can be attributed to the heterogeneity in the synaptic weights, an idea that has been discussed in \parencite{li_clustering_2003-1,li_clustering_2003,bogaard_interaction_2009}.  

For each coupling configuration considered, a network initialized in the synchronized %($\psi=0$) 
state, can persist under transient perturbations, applied as current pulses, to a subset of the cells.  Networks equipped with heterogeneous coupling (nearest neighbours or distance-dependent), initially set in a stable non-trivial solution were observed to synchronize in response to these same transient current pulses, applied to a specific number of cells. Additionally, networks with nearest neighbours coupling, initialized in a splay-state solution (number of clusters equals the number of neurons), could also switch to a different stable cluster solution, if a different subset of cells received the current pulses. Our results indicate that under low levels of ACh, the susceptibility of a pyramidal network to persist in response to incoming transient stimuli can be modulated by the changes in the coupling. This is partly consistent with the hypothesis outlined in \parencite{hasselmo_neuromodulation_1999, hasselmo_encoding_1996}, which suggests that low levels of ACh observed during memory retrieval reduces the overall responsiveness of the neurons to afferent input.

As we\textit{ increase the level of ACh in the network $(g_M: 5 \rightarrow 0)$}, the stability trends become a little complicated. Across all the coupling architectures considered, the network can desynchronize into multiple stable cluster solutions. It is generally difficult to find a consistent metric to compare the total number of distinct stable solutions among the three cases. However, we can say that when heterogeneity is introduced into the coupling architecture,  a given network can admit a larger \textit{number of clusters }($n$). Although a network with globally homogeneous coupling supports fewer distinct values of $n$ for which the corresponding $n-$cluster solutions are stable, it may still exhibit a larger number of different realizations (via groupings of the neurons) of the same $n-$cluster. This is due to the lack of constraints on the firing order of the clusters and the flexibility in which cells are recruited into each cluster, which is absent in the case of heterogeneous coupling. \textit{Under high levels of ACh}, the neural network can exhibit \textit{multistablity} with fixed \textit{synaptic weights}. This presents the network with a mechanism to respond to different inputs, which may alter the average network frequency, without modifying the strengths of the neural connectivity.   Additionally, the multistability associated with the fixed synaptic weights can also present a mechanism for the network to transition between different cluster states. This was observed in our results, through the application of transient current pulses to a subset of the cells in the network, initialized at a stable cluster solution. In the simulations, the transient current pulse could switch the network to alternate stable cluster solutions. This observation was consistent across all simulations, regardless of network sizes, $N$, and coupling configuration, $W$. These results indicate that under high levels of ACh, the network is largely susceptible to transient changes in stimuli, consistent with Class I dynamics \parencite{izhikevich_dynamical_2007,hasselmo_modes_2011,roach_acetylcholine_2019}, in contrast to the low ACh case. 
 
Traditionally, high levels of ACh observed during active waking, have been linked to the initial encoding of novel episodic memories \parencite{hasselmo_modes_2011,hasselmo_dynamics_1995,hasselmo_high_2004,hasselmo_encoding_1996}.
In our results, the multistability exhibited across all configurations of the network under high ACh ($g_M=0$) can support the rapid encoding of distinct memories without synaptic modification (learning) by presenting the hippocampal network with a predefined set of unique neural assembly patterns (cluster solutions) to choose from. 

Networks equipped with an all-to-all coupling, can resemble a network of recurrently connected pyramidal neurons in the CA3 region of the hippocampus. This region has also been linked to the efficient storage of memories, where a single CA3 pyramidal cell can be involved in distinct representations of multiple memories \parencite{rolls_theory_1996}.   When the M-current is suppressed in the presence of acetylcholine, the behaviour of CA3 is predominantly influenced by the external inputs coming in through the entorhinal cortex \parencite{hasselmo_neuromodulation_1999}. This can enhance the network's responsiveness to incoming signals, allowing it to flexibly encode new information.  In particular, this facilitates an encoding process known as \textit{pattern separation}, involving the dentate gyrus and the CA3, in which pairs of incoming stimuli with subtle differences, are encoded as two distinct representations with a low degree of similarity \parencite{madar_pattern_2019,faghihi_computational_2015}. In the context of our results, the network may initially exhibit a pattern of activity represented by a cluster solution, to encode some stimulus, `A'.  Suppose this network has encode a separate stimulus `B'  sharing subtle differences with `A'.  Their differences may be conveyed via transient current pulses received by the initial cluster solution (associated with 'A') inducing a transition into a different  $n-$ cluster solution to be associated with the incoming stimulus 'B'.  In this interpretation,  we assume that the stimuli are presented sequentially—`A' is encoded first, followed by `B'.   Both the multistability and the high susceptibility observed in our results, enable the same network to encode two distinct but similar stimuli rapidly without modifying the synaptic strengths.

In contrast, low levels of ACh observed during stages of quiet waking and slow-wave sleep in the hippocampus block further encoding of novel information into CA3, to facilitate the consolidation and retrieval of memories \parencite{hasselmo_encoding_1996}. This may be reflected in our results for the all-to-all network, through the general lack of distinct n-cluster solutions, in addition to the selective persistence under injected current pulse. Especially for networks equipped with all-to-all coupling, respond to a lesser extent to the current pulses thereby preventing further encoding from occurring.

%Other Types of solutions
Through this framework, we demonstrated differences in the neural assemblies supported by a network of hippocampal pyramidal neurons coupled via weak excitatory AMPA synapses, under varying levels of acetylcholine (ACh) and across three different coupling architectures. Overall, our results indicate that ACh likely plays a role in guiding neural assembly formation within the hippocampal network—potentially linking specific assembly patterns to distinct behavioural states, particularly those related to the encoding and consolidation, which are hypothesized to be regulated by different levels of ACh. 
% Through this framework, we examined two extreme configurations of the network coupling---one where each neuron is coupled to its nearest neighbours and the other where each neuron is coupled to every other neuron with the same strength of excitatory synapse. On the one hand, the nearest neighbours showed little to no difference under different levels of ACh beyond the stability of  the anti-phase and in-phase state. On the other hand the globally homogeneous network showed (???). It is  
The stark contrast in the extent of multistability exhibited by the network at high versus low levels of ACh qualitatively supports the findings of \parencite{roach_acetylcholine_2019}, which identified two distinct frequency-dependent coding schemes for information processing corresponding to these ACh levels. However, our analysis, primarily exploratory in nature, was restricted to a specific class of solutions in which phase differences between neural oscillators were assumed to be uniform. Consequently, each cluster contained an identical number of neurons, spatially dispersed throughout the network. More complex cluster states may arise outside this assumption, and exploring them could offer a more complete picture of the role of ACh in shaping hippocampal network dynamics. For example, in \parencite{li_clustering_2003-1}, the authors studied model independent cluster solutions for a ring network equipped with asymmetrical coupling strengths which added more heterogeneity in the network structure, providing a more realistic view on the assembly formation.  In \parencite{ryu_spatially_2021}, the focus was on spatially compact assemblies in inhibitory networks, where the cells belonging to the same cluster were also located near one another one on the $1-$dimensional ring. %However this was suggested as a mechanism appropriate for a network of inhibitory neurons as opposed to one which is excitatory. 
In this context, it would be worthwhile to explore such solutions for a network of GABAergic interneurons located within the CA3 region of the hippocampus, which also contain an M-current. Additionally, our analysis also assumes specific periodic conditions which restrict the model neurons to the $1-$dimensional ring. Another possible next step could involve extending our analysis to $2-$dimensional networks, similar to the approach taken in \parencite{miller_patterns_2022}, but in an acetylcholine-specific context. 

% Other effects of acetylcholine in our model
One of the limitations in our neuron model, is the assumption that the restriction of the acetylcholine effect at the cellular level is to the M-current. Acetylcholine also suppresses a calcium activated potassium current, $I_{AHP}$ \footnote{AHP here stands for afterhyperpolarization.} \parencite{madison_voltage_1987} and the leak current, $I_L$ \parencite{stiefel_cholinergic_2008}. Since neither current contribute to the switch in the neural class excitability, they were not considered in our analysis. In \parencite{al-darabsah_m-current_2021}, it is shown that the simultaneous down-regulation of the leak and M-current would increase the value of $g_M$ at which the excitability switch will occur. In our analysis, this, potentially creates a broader range of $g_M$, for which the corresponding network can exhibit multistability and an increased sensitivity to external inputs. We leave this exploration along with the effect of $I_{AHP}$ current for future work.  Our analysis did not address the biophysical implications of intermediate ACh levels ($0<g_M<5$).  Previous studies \parencite{roach_formation_2015}, suggest that at these intermediate levels, the hippocampal network can exhibit dynamics sensitive to input heterogeneities and may integrate both new and existing activity patterns. 

% Phase model limitations and future directions. 
One caveat of a phase model approximation is that it requires that the neurons are intrinsically oscillating and coupled via sufficiently weak synapses. To satisfy this condition, we incorporated a constant external current, $I_{app}$, into the model neuron, that was large enough to ensure intrinsic oscillations for all values of $g_M$ between $0$ and $5$. However, this approach resulted in neurons firing at high frequencies, exceeding  $500$ Hz when $g_M$ was set to $0$, which is beyond the typical biophysical range associated with observed brain rhythms. To model more realistic firing rates, one possible approach would be to restrict the range of $g_M$ and choose a smaller applied current. Alternatively,  the applied current could be adjusted depending on the $g_M$ value to maintain a fixed firing frequency or include inhibitory neurons in the network. The latter approach has been considered in \parencite{roach_acetylcholine_2019}, with a computational model. While the assumption of weak coupling is not entirely biophysically unrealistic, a valuable direction could be to examine the persistence of these model predictions under stronger coupling conditions, particularly comparing behaviours under low versus high ACh levels. Finally, applying phase model analysis to inhibitory interneuron networks incorporating the M-current could reveal whether their synchrony dynamics align with the numerical findings observed in pairs of inhibitory neurons, as reported in  \parencite{al-darabsah_m-current_2021}. A further extension could then involve deriving a phase model for a mixed excitatory-inhibitory network and analyzing the corresponding cluster solutions \parencite{ermentrout_mathematical_2010}, providing a more biophysically realistic model for the hippocampus.  

In spite of its limitations, the phase model approach remains a powerful and insightful tool for analyzing synchrony in neuronal networks. It enables the reduction of high dimensional biophysical systems to simpler mathematical models, facilitating a systematic approach to the investigation of neural synchrony, under  varying physiological conditions, such as different levels of acetylcholine.

% \section{References}
% \bibliographystyle{abbrv}
% \bibliographystyle{apalike}
%\bibliographystyel{}

\printbibliography

%\bibliographystyle{apa} % should get .bst file from publisher
%\bibliography{references,other}% multiple bib file

\end{document}